\documentclass[graybox]{svmult}

\usepackage{type1cm}
\usepackage{makeidx}
\usepackage{graphicx}
\usepackage{multicol}
\usepackage[bottom]{footmisc}
\usepackage{newtxtext}
\usepackage{newtxmath}

\usepackage{color}
\usepackage{amsmath,amsfonts}
\usepackage{graphicx}  
\usepackage{epsfig,epstopdf}
\usepackage{multirow}

\spnewtheorem{remark}{Remark}{\bfseries}{\rmfamily}

\spnewtheorem{example}{Example}{\bfseries}{\rmfamily}

\makeindex 

\begin{document}

\title*{Virtual Rings on Highways: Traffic Control by Connected Automated Vehicles}
\author{Tam\'as G. Moln\'ar, Michael Hopka, Devesh Upadhyay, Michiel Van Nieuwstadt, and G\'abor Orosz}
\authorrunning{T. G. Moln\'ar, M. Hopka, D. Upadhyay, M. Van Nieuwstadt, and G. Orosz}
\institute{Tam\'as G. Moln\'ar \at Department of Mechanical and Civil Engineering, California Institute of Technology, Pasadena, CA 91125, USA, \email{tmolnar@caltech.edu}
\and Michael Hopka, Devesh Upadhyay, and Michiel Van Nieuwstadt \at Ford Research and Innovation Center, Dearborn, MI 48124, USA, \email{mhopka@ford.com, dupadhya@ford.com, mvannie1@ford.com}
\and G\'abor Orosz \at Department of Mechanical Engineering and Department of Civil and Environmental Engineering, University of Michigan, Ann Arbor, MI 48109, USA, \email{orosz@umich.edu}}

\maketitle

\abstract{
This work gives introduction to traffic control by connected automated vehicles.
The influence of vehicle control on vehicular traffic and traffic control strategies are discussed and compared.
It is highlighted that vehicle-to-everything connectivity allows connected automated vehicles to access the state of the traffic behind them such that feedback can be utilized to mitigate evolving congestions.
Numerical simulations demonstrate that such connectivity-based traffic control is beneficial for smoothness and energy efficiency of highway traffic.
The dynamics and stability of traffic flow, under the proposed controllers, are analyzed in detail to construct stability charts that guide the selection of stabilizing control gains.
}

\begin{figure}[!t]
\begin{center}
\includegraphics[scale=.67]{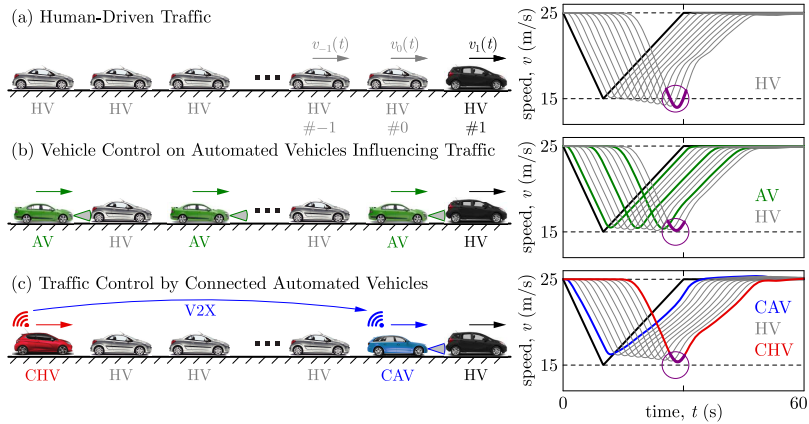}
\caption{
Dynamics of traffic flows.
(a) A chain of human-driven vehicles (HVs).
If the lead vehicle brakes, the vehicle chain amplifies it and the tail vehicle reduces its speed more than the lead.
(b) Vehicle control on large penetration of automated vehicles (AVs) which influences traffic.
The tail vehicle reduces its speed about as much as the lead.
(c) Traffic control with a single connected automated vehicle (CAV) where the tail vehicle is a connected human-driven vehicle (CHV).
The tail vehicle reduces its speed less than the lead.
}
\label{fig:concept}
\end{center}
\end{figure}

\section{Introduction}
\label{sec:intro}

Traffic congestion is a major factor in reducing the efficiency of road transportation, as it increases travel times, energy consumption of vehicles and air pollution.
Human-driven traffic often suffers from the onset of stop-and-go traffic jams on highways.
These may occur without any incident and may be triggered by the human driving behavior~\cite{Orosz2009}.
This is illustrated in Fig.~\ref{fig:concept}(a), where a chain of human-driven vehicles (HVs) is simulated on a single lane (with details given later in the paper).
When the lead vehicle (black) decelerates, the following human drivers (gray) tend to overreact and reduce their speed more: they get slower the farther they are behind the lead vehicle.
This, so-called string unstable driving behavior is typical in human-driven traffic~\cite{Molnar2021trc}.
It eventually leads to low speeds and stop-and-go motion on highways, giving birth to a traffic congestion~\cite{Orosz2010}.

This chapter is dedicated to strategies that allow prevention or mitigation of such driving behavior-induced traffic congestions.
Alleviating traffic jams has substantial benefits for each individual vehicle participating in traffic: it improves their travel time and energy consumption.
The key factor to obtain these benefits is to ensure that vehicles drive as smoothly as possible without major deceleration.
Smoothness is often analyzed via the notion of string stability~\cite{Besselink2017, Feng2019, Ploeg2014a, Swaroop1996}, which also has been used recently to evaluate commercial cruise control systems~\cite{Ciuffo2021, Gunter2020arecommercially, Gunter2020model}.
Smooth driving can be achieved either at the level of individual vehicles by {\em vehicle control} or at large-scale traffic level by {\em traffic control}~\cite{DelleMonache2019}.

Vehicle control can be realized on automated vehicles (AVs), with various levels of automation.
Control strategies include adaptive cruise control (ACC) for individual AVs~\cite{Xiao2010}, connected cruise control (CCC) for connected automated vehicles~\cite{Orosz2016}, and cooperative adaptive cruise control (CACC) for vehicle platoons~\cite{Wang2018b}.
Controlling vehicles to drive smoothly achieves significant benefits in energy efficiency, safety and passenger comfort~\cite{Dollar2018, Ge2018, Zheng2020}.
While vehicle control guarantees benefits for the individual AVs only, a sufficient number of well-designed AVs may positively influence traffic at large and mitigate traffic jams due to their smooth driving behavior~\cite{Ard2020, Arem2006, Cui2017, Spiliopoulou2018}.
The influence of vehicle control on traffic is demonstrated in Fig.~\ref{fig:concept}(b), where AVs (green) are mixed with HVs (gray) in the traffic flow.
As the AVs drive smoothly and reduce their speed less than the vehicle they follow, the traffic jam is mitigated.
This way, vehicle control also controls traffic indirectly, however, there is no direct measurement or feedback of the traffic state.

Traffic control, on the other hand, directly uses the state of the traffic flow as feedback.
Traffic state estimation may rely on data from loop detectors or cameras at fixed locations~\cite{Mehran2011, Orosz2010, Yu2019a} or from vehicles traveling in the traffic flow~\cite{Bekiaris-Liberis2017, Chu2016, Chu2019, DelleMonache2019a}.
State estimators may involve Kalman filtering techniques~\cite{Herrera2010a, Work2008a, Work2009} and data fusion~\cite{Duret2017, Yuan2014, Yuan2012}.
Via traffic state feedback, traffic control aims to maximize the benefits of large-scale traffic, which eventually leads to benefits for individual vehicles too.
Research in traffic control has seen a surge in recent years~\cite{Siri2021}, extending from classical traffic control~\cite{DelleMonache2019} to reinforcement learning~\cite{Kreidlieh2018, Wu2018}.
Traffic control has been approached from so-called Eulerian and Lagrangian perspectives~\cite{Laval2013}.
In Eulerian traffic control, traffic flow is regulated at specific locations along the highway based on data about the traffic further down the road~\cite{Bekiaris-Liberis2020, Yu2018}.
Traffic control measures include speed limits~\cite{Karafyllis2018} and ramp metering~\cite{Koehler2016, Pasquale2018, Yu2019, Yu2021}.
In Lagrangian traffic control, traffic is regulated by the driving behavior of certain vehicles in the flow~\cite{Cicic2018, Piacentini2019}, which directly take into account how they affect traffic, and utilize the traffic state as feedback in their controllers.

Since Lagrangian traffic controllers have been enabled by vehicle automation, they have appeared more recently than Eulerian traffic control, and their literature is still less extensive.
Meanwhile, technological advancement in the field of automated vehicles keeps opening new possibilities for traffic control.
In particular, the occurrence of connected automated vehicles (CAVs) has significant potential that has not yet been addressed by the literature.
Therefore, this chapter is devoted to establishing the notion of Lagrangian traffic control by connected automated vehicles, and highlighting its potential for mitigating traffic congestions.

Namely, automated vehicles must monitor the state of the traffic behind them while traveling on highways in order to achieve Lagrangian control of the traffic behind.
Monitoring traffic can be realized by vehicle-to-everything (V2X) connectivity, which becomes the enabling technology of traffic control by connected automated vehicles.
A CAV, communicating with vehicles behind it, receives traffic state information that can be used as feedback~\cite{Chen2021a}.
This approach is demonstrated in Fig.~\ref{fig:concept}(c), where a CAV communicates with and responds to a connected human-driven vehicle (CHV) behind it.
Information from the CHV (red) allows the CAV (blue) to mitigate the traffic jam efficiently: the entire vehicle chain reduces its speed less than for the case with a large number of AVs in Fig.~\ref{fig:concept}(b).
This demonstrates the large potential of CAVs in traffic control that is yet to be exploited.

The feedback loop in Fig.~\ref{fig:concept}(c) is similar to when vehicles travel on a ring road: they respond to each other in a loop, the first vehicle responding to the last one.
Ring configurations have rich and interesting dynamics~\cite{Giammarino2019, Orosz2010, Allwoerden2021, Wang2020, Zheng2020} and there have even been successful implementations of vehicle controllers on ring roads to mitigate traffic~\cite{Stern2018}.
While the road geometry may not be a ring in practice, the response of vehicles in Lagrangian traffic control has a similar structure, which we refer to as virtual rings.
The concept of virtual rings have been introduced in the experiments of~\cite{Avedisov2018} where three vehicles performed car following on a straight highway, with the first vehicle responding to the last one.
This concept was further discussed in the context of traffic control in~\cite{Avedisov2021, Molnar2020cdc}, and also utilized in~\cite{Wang2020c} where the so-called leading cruise control was introduced for high-connectivity penetration scenarios.

Here we further elaborate the concept of virtual rings and Lagrangian traffic control by CAVs.
As opposed to previous works~\cite{Molnar2020cdc, Wang2020c}, we put more emphasis on the stability analysis of traffic flow, focus on low-automation and low-connectivity scenarios, and incorporate time delays associated with the response of vehicles, their controllers and the human drivers.
We first introduce strategies for vehicle and traffic control, then highlight the benefits of connectivity-enabled feedback loops via simulations, and finally give a comprehensive stability analysis.

\section{Traffic Control by Connected Automated Vehicles}

We consider traffic regulation by means of controlling the motions of automated vehicles (AVs) and connected automated vehicles (CAVs) traveling in the traffic flow.
Since the driving behavior of these AVs and CAVs affects the motion of other vehicles behind them, including human-driven vehicles (HVs) and connected human-driven vehicles (CHVs), they regulate the traffic behind.

First, we discuss strategies where AVs and CAVs respond to vehicles ahead of them only.
In this case, AVs and CAVs indirectly control the traffic behind them, without feedback of the traffic state.
We call this case as {\em vehicle control influencing traffic}.
Then we propose strategies where CAVs respond to vehicles behind them as well.
This allows CAVs to directly control traffic by using the traffic state behind as feedback.
We refer to this case as {\em traffic control}.

We consider single lane traffic for simplicity of exposition.
The results could be extended to multiple lanes by incorporating the cross-lane dynamics.
While lane changes and overtaking requires further investigation and extensive future work, we remark that traffic controllers realized on CAVs consider the interest of the vehicles behind them, which makes it less likely for those vehicles to overtake the CAVs.

\subsection{Simplified Models for Longitudinal Vehicle and Traffic Dynamics}

Consider the scenario in Fig.~\ref{fig:concept}, in which vehicles follow each other on a single lane of a straight road.
We number the vehicles with indices increasing in the direction of motion.
We denote the set of all vehicle indices by $\mathcal{N}$ that comprises of the indices ${\mathcal{N} = \{ \mathcal{N}_{\rm HV}, \mathcal{N}_{\rm CHV}, \mathcal{N}_{\rm AV}, \mathcal{N}_{\rm CAV} \}}$ representing four vehicle types defined previously.
We distinguish a so-called {\em ego vehicle} of interest by index $0$.
We denote the length of vehicle $n$ by $l_{n}$, the position of its rear bumper by $s_{n}$ and its speed by $v_{n}$, ${n \in \mathcal{N}}$.

We model the motion of vehicle $n$ by a delayed double integrator with saturation:
\begin{align}
\begin{split}
\dot{s}_{n}(t) & = v_{n}(t) \,, \\
\dot{v}_{n}(t) & = {\rm sat} (u_{n}(t-\tau_{n})) \,,
\end{split}
\label{eq:dyn}
\end{align}
${\forall n \in \mathcal{N}}$, where $u_{n}$ is the desired acceleration of vehicle $n$, selected by the human driver for HVs and CHVs (${n \in \{\mathcal{N}_{\rm HV}, \mathcal{N}_{\rm CHV}\}}$) and prescribed by the longitudinal controller of AVs and CAVs (${n \in \{\mathcal{N}_{\rm AV}, \mathcal{N}_{\rm CAV}\}}$).
We assume that each vehicle realizes the desired acceleration by the help of human action or low-level controllers, respectively, unless this acceleration is above the acceleration capability $a_{\rm max}$ or below the braking limit $-a_{\rm min}$ of the vehicle.
This is captured by the saturation function:
\begin{equation}
{\rm sat}(u) = {\rm min} \{ {\rm max} \{ -a_{\rm min},u \}, a_{\rm max} \} \,.
\end{equation}
Furthermore, we incorporate the time delay $\tau_{n}$ into the model, that involves the response time of the vehicle, as well as the driver reaction time for HVs and CHVs, and feedback delays for AVs and CAVs.
For simplicity, we assume identical delays for HVs and CHVs throughout this study with ${\tau_{n} = \tau}$, ${n \in \{\mathcal{N}_{\rm HV}, \mathcal{N}_{\rm CHV}\}}$, and we distinguish the delay of the ego vehicle $0$ by the notation ${\sigma = \tau_{0}}$.

To capture human driver behavior, we use simple car-following models.
Namely, our {\em human driver model (HDM)} assumes that vehicle $n$ responds to vehicle ${n+1}$ ahead considering the headway (range) ${h_{n}=s_{n+1}-s_{n}-l_{n}}$ and the speed difference (range rate) ${\dot{h}_{n} = v_{n+1}-v_{n}}$ between them:
\begin{equation}
u_{n} = f_{\rm H}(
h_{n},
\dot{h}_{n},
v_{n}
) \,,
\label{eq:HDM}
\end{equation}
where the specific expression of $f_{\rm H}$ depends on the choice of the model.
Examples for HDM include the optimal velocity model (OVM)~\cite{Bando1998} and the intelligent driver model (IDM)~\cite{Treiber2000}, which were shown to capture human driver behavior observed in experimental data~\cite{Avedisov2018, Dollar2021}.
For simplicity of exposition, throughout this paper we assume that human drivers are identical in their driving behaviors and $f_{\rm H}$ is the same for each ${n \in \{\mathcal{N}_{\rm HV}, \mathcal{N}_{\rm CHV}\}}$.

\begin{example}\label{exmp:HDM}
For the numerical examples of this paper, we use the optimal velocity model as human driver model:
\begin{equation}
f_{\rm H}(h_{n}, \dot{h}_{n}, v_{n})
= \alpha_{\rm H} \big( V_{\rm H}(h_{n}) - v_{n} \big)
+ \beta_{\rm H} \dot{h}_{n} \,.
\label{eq:OVM}
\end{equation}
Here ${\alpha_{\rm H}}$ and ${\beta_{\rm H}}$ are parameters describing the human driver.
The second term with coefficient $\beta_{\rm H}$ captures the response of human drivers to the speed difference ${\dot{h}_{n} = v_{n+1}-v_{n}}$ relative to the vehicle ahead. 
The first term with parameter $\alpha_{\rm H}$ characterizes the response to the headway ${h_{n}=s_{n+1}-s_{n}-l_{n}}$ measured from the vehicle ahead.
We assume that human drivers track a headway-dependent desired speed given by the range policy:
\begin{equation}
V_{\rm H}(h) = {\rm min}\{{\rm max}\{0,F_{\rm H}(h)\},v_{\rm max}\} \,.
\label{eq:rangepolicy}
\end{equation}
This desired speed is nonnegative and it saturates at the speed limit $v_{\rm max}$.
The speed increases strictly monotonically between $0$ and $v_{\rm max}$ at a rate defined by $F_{\rm H}$ that is a function of the headway $h$.
\end{example}

\subsection{Vehicle Control Influencing Traffic}

\begin{figure}[!t]
\begin{center}
\includegraphics[scale=.67]{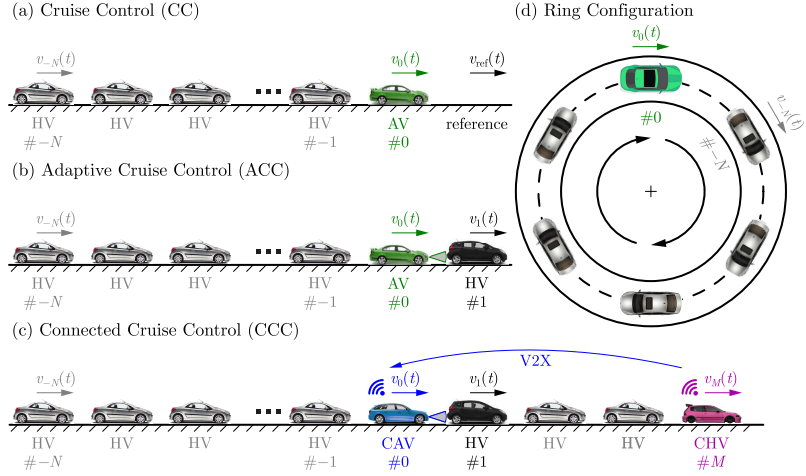}
\caption{
Vehicle control strategies where the driving behavior of (connected) automated vehicles influences the traffic behind them, including (a) cruise control, (b) adaptive cruise control, (c) connected cruise control, and (d) these strategies realized on a ring road.
}
\label{fig:openloop}
\end{center}
\end{figure}

Now we consider vehicle control for (connected) automated vehicles where AVs and CAVs influence the traffic behind them without explicitly responding to its state.
Consider the scenario in Fig.~\ref{fig:openloop}(a)-(c).
We assume that the ego vehicle (vehicle  $0$) is automated (an AV in Fig.~\ref{fig:openloop}(a)-(b) and a CAV in Fig.~\ref{fig:openloop}(c)), and it is followed by $N$ human-driven vehicles (HVs) constituting the traffic influenced by vehicle control.

Our goal is to design a control input $u_{0}$ for the ego vehicle based on the traffic ahead and observe its effects on the behavior of the traffic behind.
If traffic conditions are free-flowing and there are no vehicles ahead of the AV such as in Fig.~\ref{fig:openloop}(a), then the AV may use {\em cruise control (CC)} to track a reference speed $v_{\rm ref}(t)$ such that its speed $v_{0}(t)$ approaches $v_{\rm ref}(t)$: 
\begin{equation}
u_{0} = f_{\text{CC}}(
v_{0},
\hspace{-1mm}\underbrace{v_{\rm ref}}_{\text{reference}}\hspace{-1mm}
) \,.
\label{eq:CC_general}
\end{equation}
The specific form of $f_{\text{CC}}$ depends on the controller type.
The reference speed $v_{\rm ref}(t)$ may be constant, or time-varying, or it may even depend on the position $s_{0}(t)$ --- an example for the latter case is when vehicles optimize their set speed based on road elevation~\cite{He2016}.

If the traffic is more dense, the AV may need to respond to the vehicle ahead; see Fig.~\ref{fig:openloop}(b).
This can be achieved by sensing the position and speed of the vehicle ahead via on-board sensors (radar, lidar, camera or ultrasonics) and using {\em adaptive cruise control (ACC)} that takes into account the positions and speeds of the AV and the vehicle ahead:
\begin{equation}
u_{0} = f_{\text{ACC}}(
s_{0},v_{0},
\hspace{-3mm}\underbrace{s_{1},v_{1}}_{\text{vehicle ahead}}\hspace{-3mm}
) \,.
\label{eq:ACC_general}
\end{equation}

Finally, if the ego vehicle is equipped with a communication device, i.e., it is a CAV, then it may respond to connected human-driven vehicles (CHVs) farther ahead, as shown by Fig.~\ref{fig:openloop}(c).
This leads to {\em connected cruise control~(CCC)} that may potentially take into account the positions and speeds of $M$ vehicles ahead:
\begin{equation}
u_{0} = f_{\text{CCC}}(
s_{0},v_{0},
\underbrace{s_{1},v_{1}, \ldots, s_{M},v_{M}}_{\text{multiple vehicles ahead}}
) \,.
\label{eq:CCC_general}
\end{equation}
If a vehicle ahead is not connected to or sensed by the CAV, its state can be omitted from $f_{\text{CCC}}$.
An aggregated response to multiple vehicles ahead facilitates smooth driving by making the CAV more clairvoyant about upcoming changes in traffic conditions~\cite{Ge2018}.

\begin{example}\label{exmp:ACC}
Throughout this paper, we consider the following simple examples as controllers for AVs and CAVs that influence the traffic behind.
The simplest CC strategy is a proportional controller for speed tracking with a control gain $\beta$:
\begin{equation}
f_{\text{CC}}(v_{0}, v_{\rm ref}) = \beta \big( v_{\rm ref} - v_{0} \big) \,.
\label{eq:CC}
\end{equation}
Indeed, this law could be replaced by other more sophisticated controllers.

For ACC, we consider the following control law:
\begin{equation}
f_{\text{ACC}}(s_{0}, v_{0}, s_{1}, v_{1}) = \alpha \big( V(s_{1} - s_{0} - l_{0}) - v_{0} \big) + \beta \big( W(v_{1}) - v_{0} \big) \,.
\label{eq:ACC}
\end{equation}
This is an analog to the optimal velocity model~(\ref{eq:OVM}), with the difference that the control gains $\alpha$, $\beta$ and the range policy $V$ can be selected as control design parameters.
For $V$ we keep the form~(\ref{eq:rangepolicy}) and prescribe the desired speed-headway relationship via $F$.
Furthermore, in~(\ref{eq:ACC}) we include the speed policy $W$ that allows the AV to follow the speed $v_{1}$ of the vehicle ahead or the speed limit $v_{\rm max}$, whichever is smaller:
\begin{equation}
W(v_{1}) = {\rm min}\{v_{1},v_{\rm max}\} \,.
\end{equation}

Finally, for CCC we extend the controller~(\ref{eq:ACC}) with response to the speed of multiple vehicles ahead~\cite{Zhang2016}:
\begin{equation}
f_{\text{CCC}}(s_{0}, v_{0}, s_{1}, v_{1}, \ldots, s_{M}, v_{M}) = \alpha \big( V(s_{1} - s_{0} - l_{0}) - v_{0} \big) + \sum_{m=1}^{M} \beta_{m} \big( W(v_{m}) - v_{0} \big) \,,
\label{eq:CCC}
\end{equation}
where $\beta_{m}$ is the control gain associated with the speed $v_{m}$ of vehicle $m$ ahead.
If some vehicle $m$ is not connected and cannot be perceived by the CAV, we omit the response to it by taking ${\beta_{m}=0}$.
\end{example}

\subsection{Traffic Control}

\begin{figure}[!t]
\begin{center}
\includegraphics[scale=.67]{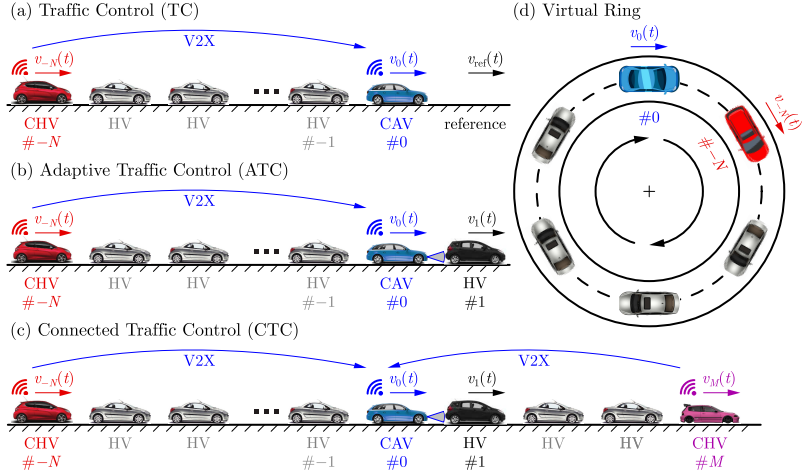}
\caption{
Traffic control strategies executed by connected automated vehicles responding to the traffic behind, including (a) traffic control, (b) adaptive traffic control, (c) connected traffic control, and (d) the virtual ring associated with these traffic control strategies.
}
\label{fig:closedloop}
\end{center}
\end{figure}

Now consider the scenario illustrated in Fig.~\ref{fig:closedloop}(a)-(c), where the ego vehicle is a CAV and at least one vehicle behind it is connected, e.g., a CHV.
Once the CAV syncs with the CHV, it can actively regulate the traffic behind, since it uses traffic state feedback when responding to the CHV.
When there are no vehicles ahead of the CAV to respond to, such as in Fig.~\ref{fig:closedloop}(a), the cruise control~(\ref{eq:CC_general}) can be extended to {\em traffic control (TC)} in response to the positions and speeds of the connected vehicles behind the ego CAV:
\begin{equation}
u_{0} = f_{\text{TC}}(
\underbrace{s_{-N},v_{-N}, \ldots,s_{-1},v_{-1}}_{\text{traffic behind}},
s_{0},v_{0},
\hspace{-1mm}\underbrace{v_{\rm ref}}_{\text{reference}}\hspace{-1mm}
) \,.
\label{eq:TC_general}
\end{equation}
The quantities ${s_{-N},v_{-N}, \ldots,s_{-1},v_{-1}}$ represent the state of the traffic behind the CAV.
We assume that not all vehicles behind the ego CAV are connected, thus, only some of these states (corresponding to connected vehicles) may be available to the CAV and affect $u_{0}$.
Accordingly, $-N$ denotes the index of the farthest connected vehicle behind communicating with the ego CAV.
This scenario is ideal for traffic control in the sense that the CAV does not have to respond to the traffic ahead in order to minimize collision risks.
Such large control freedom enables the CAV to stabilize long vehicle chains behind it.

However, when the CAV is traveling in denser traffic environments like the one in Fig.~\ref{fig:closedloop}(b), it has to respond to the vehicle ahead and has less freedom to regulate the traffic behind it.
In such a setup, the adaptive cruise control~(\ref{eq:ACC_general}) can be extended with feedback from the traffic behind to {\em adaptive traffic control (ATC)}:
\begin{equation}
u_{0} = f_{\text{ATC}}(
\underbrace{s_{-N},v_{-N}, \ldots,s_{-1},v_{-1}}_{\text{traffic behind}},
s_{0},v_{0},
\hspace{-3mm}\underbrace{s_{1},v_{1}}_{\text{vehicle ahead}}\hspace{-3mm}
) \,.
\label{eq:ATC_general}
\end{equation}

Finally, if vehicles ahead of the CAV are also connected, as illustrated by Fig.~\ref{fig:closedloop}(c), we may establish {\em connected traffic control (CTC)} that responds to multiple vehicles ahead while also taking into account its effect on the traffic behind:
\begin{equation}
u_{0} = f_{\text{CTC}}(
\underbrace{s_{-N},v_{-N}, \ldots,s_{-1},v_{-1}}_{\text{traffic behind}},
s_{0},v_{0},
\underbrace{s_{1},v_{1}, \ldots, s_{M},v_{M}}_{\text{multiple vehicles ahead}}) \,.
\label{eq:CTC_general}
\end{equation}
\begin{remark}\textbf{-- Connectivity and automation.}
We remark that although only CTC includes the word "connected" in its name, TC and ATC also rely on connectivity to obtain information from the traffic behind (unless ${N=1}$ and the vehicle behind the AV is detected by on-board sensors).
At the same time, automation is not required by any of these strategies for vehicles other than the ego vehicle.
Yet, if other connected vehicles happen to possess sufficient levels of automation (they are CAVs), that opens the possibility of coordinating CAVs for traffic control~\cite{Mahbub2021, Mahbub2021a, Piacentini2018}.
In this paper, we will not discuss details about such coordination.
\end{remark}

\begin{example}\label{exmp:ATC}
Analogously to the CCC strategy in~(\ref{eq:CCC}), response to the traffic behind can be achieved from the feedback of the speed of each connected vehicle behind.
This leads to the TC algorithm:
\begin{equation}
u_{0} = \beta \big( v_{\rm ref} - v_{0} \big) + \sum_{n=-N}^{-1} \beta_{n} \big( W(v_{n}) - v_{0} \big) \,,
\label{eq:TC_multi}
\end{equation}
cf.~(\ref{eq:CC}).
Again, ${\beta_{n}=0}$ is considered if the CAV does not perceive a vehicle due to the lack of sensors or connectivity.

The ACC strategy~(\ref{eq:ACC}) can also be extended to ATC with a feedback term from the traffic behind:
\begin{equation}
u_{0} = \alpha \big( V(s_{1}-s_{0}) - v_{0} \big) + \beta \big( W(v_{1}) - v_{0} \big) + \sum_{n=-N}^{-1} \beta_{n} \big( W(v_{n}) - v_{0} \big) \,.
\label{eq:ATC_multi}
\end{equation}
In what follows, we focus on scenarios with lean penetration of connectivity.
If only a single vehicle is connected to the CAV, we obtain
\begin{equation}
u_{0} = \alpha \big( V(s_{1}-s_{0}) - v_{0} \big) + \beta \big( W(v_{1}) - v_{0} \big) + \beta_{\rm B} (W(v_{-N}) - v_{0}) \,,
\label{eq:ATC}
\end{equation}
where ${\beta_{\rm B} = \beta_{-N}}$ is the control gain related to the traffic behind.

Finally, the CCC~(\ref{eq:CCC}) can be extended to CTC according to
\begin{equation}
u_{0} = \alpha \big( V(s_{1}-s_{0}) - v_{0} \big) + \sum_{m=1}^{M} \beta_{m} \big( W(v_{m}) - v_{0} \big) + \sum_{n=-N}^{-1} \beta_{n} \big( W(v_{n}) - v_{0} \big) \,.
\label{eq:CTC_multi}
\end{equation}
\end{example}

\begin{figure}[!t]
\begin{center}
\includegraphics[scale=.67]{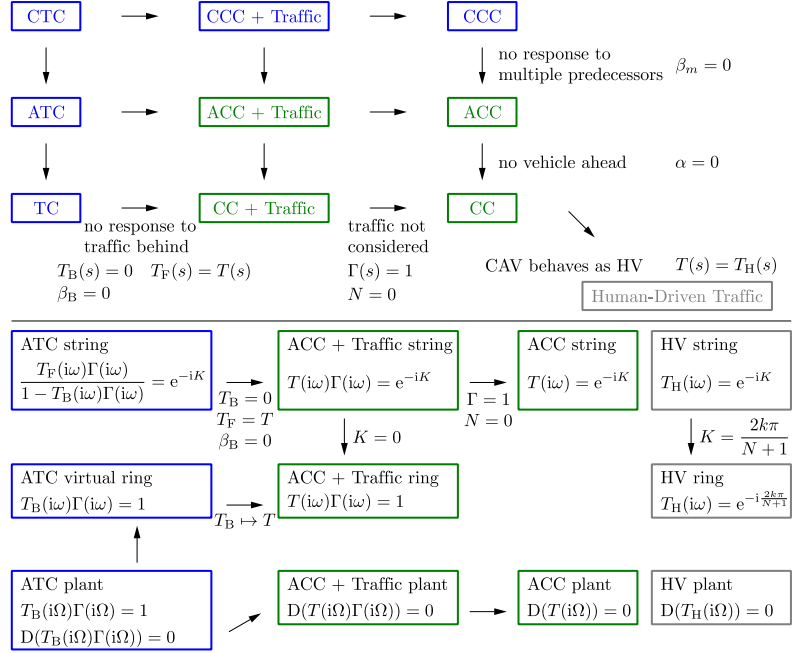}
\caption{
Relationships between control strategies (top) and the associated stability conditions (bottom).}
\label{fig:interconnections}
\end{center}
\end{figure}

\begin{remark}\textbf{-- Relationships between control strategies.}
All of the above vehicle and traffic controllers can be regarded as special cases of CTC.
Ultimately, the choice of controller depends on the available information: whether there are vehicles ahead and behind the ego CAV that are connected to it.
The various scenarios are summarized at the top of Fig.~\ref{fig:interconnections}.
When the ego CAV does not respond to multiple vehicles ahead, the CTC and CCC strategies reduce to ATC and ACC.
When there are no vehicles ahead of the CAV at all, ATC and ACC reduce to TC and CC.
When the CAV does not respond to the traffic behind, the CTC, ATC and TC strategies reduce to a CAV executing CCC, ACC and CC, respectively, that influences the following human-driven traffic.
Taking ${N=0}$ we can disregard the influence of the CAV on the traffic behind, leading to CCC, ACC and CC controllers without considering traffic environment.
Finally, if the CAV behaves as a human driver, we recover the case of human-driven traffic.
We will also see these interconnections in the formulas describing the stability of traffic flow.
\end{remark}

\begin{remark}\textbf{-- Rings and virtual rings.}
Longitudinal controllers of AVs and CAVs are often analyzed on ring roads, as illustrated by Fig.~\ref{fig:openloop}(d).
Mathematically, the ring configuration has an additional periodic boundary condition.
For example, if there are $N+1$ vehicles (indexed from $-N$ to $0$) on a ring, we have:
\begin{align}
\begin{split}
s_{-N}(t) & = s_{1}(t) \,, \\
v_{-N}(t) & = v_{1}(t) \,.
\end{split}
\label{eq:periodicBC}
\end{align}
This is a useful property for the stability analysis of traffic, as discussed below.
Apart from analysis~\cite{Giammarino2019, Orosz2010, Allwoerden2021, Wang2020, Zheng2020}, ring roads have been used in experiments, including successful mitigation of traffic jams~\cite{Stern2018}.

The ring configuration is important conceptually for traffic control that involves feedback of the traffic behind.
Namely, the structure of which vehicles respond to which other vehicles includes a ring (a loop), see Fig.~\ref{fig:closedloop}(d), without physically traveling on a ring road.
Therefore, we refer to the structure of traffic control as {\em virtual ring}.
The concept of virtual ring appeared in the experiments of~\cite{Avedisov2018} and the analysis of~\cite{Avedisov2021, Molnar2020cdc}.
Below we build on these works and give a more detailed analysis.

We also remark that traffic control on single-lane ring roads has a significant advantage compared to single-lane straight roads: the CAV can affect the equilibrium speed of the ring and can decide to travel slower than the set speed of the vehicle ahead.
Once the CAV slows down, all vehicles including the vehicle ahead will eventually travel slower on the ring.
In contrast, if the CAV decides to travel slower on a straight road, it will fall behind the vehicle ahead and break up the car-following scenario.
Hence the virtual ring is more appealing for highway applications than considering a physical ring road.
\end{remark}

\begin{remark}\textbf{-- Scenario in the rest of the paper.}
In what follows, we analyze the dynamics and stability of traffic influenced by vehicle control and traffic control.
For simplicity, we consider ACC, like in Fig.~\ref{fig:openloop}(b), and ATC with a single CHV behind the ego CAV, like in Fig.~\ref{fig:closedloop}(b).
This restricts us to lean penetration of connectivity and we can compare how vehicle control without connectivity influences traffic to traffic control allowed by connectivity.
We use human driver model~(\ref{eq:OVM}), ACC~(\ref{eq:ACC}) and ATC~(\ref{eq:ATC}) as example.
Vehicle $1$ is the lead vehicle, vehicle $0$ is the ego CAV, and vehicles $-1, \ldots, -N$ are HVs with $-N$ being connected in the ATC setup. 
Furthermore, we analyze ACC on a ring road and relate it to the virtual ring of ATC.
\end{remark}

\section{Benefits of Connectivity}

Now we demonstrate that information from connectivity is highly beneficial for traffic control.
We compare the ACC and ATC setups by investigating the traffic behavior through numerical simulations.
We also evaluate the energy consumption of vehicles and show that utilizing connectivity leads to energy savings.
The parameters used for numerical case studies throughout this paper are listed in Table~\ref{tab:parameters} in the Appendix.

\begin{figure}[!t]
\begin{center}
\includegraphics[scale=.67]{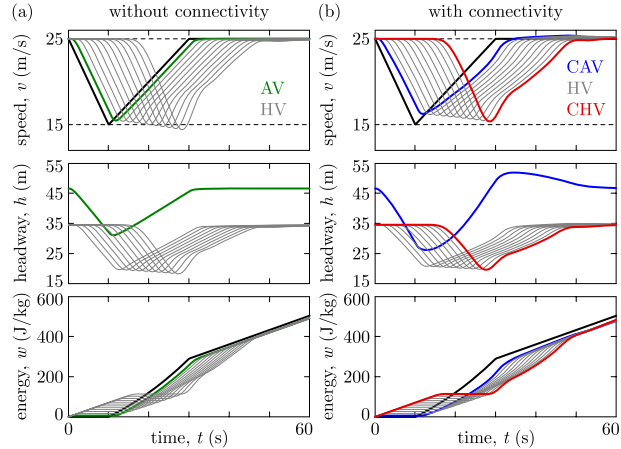}
\caption{
Traffic influenced by vehicle control without connectivity using adaptive cruise control (left) and regulated ty traffic control with connectivity using adaptive traffic control (right).}
\label{fig:simulation}
\end{center}
\end{figure}

\subsection{Simulation Results}

First, we study a baseline scenario without automation.
We simulate~(\ref{eq:dyn}, \ref{eq:HDM}) for human driver model~(\ref{eq:OVM}) in Example~\ref{exmp:HDM} with typical human driver parameters selected from~\cite{Avedisov2018}.
The simulation results are shown in Fig.~\ref{fig:concept}(a).
In the simulation, a lead vehicle (black) sequentially decelerates, accelerates and cruises at constant speed, followed by 11 HVs (gray).
The following HVs tend to reduce their speed more the farther they are from the lead vehicle as speed perturbations amplify along the vehicle chain.
Consequently, the tail vehicle brakes noticeably more than the lead (purple highlight).
This is called string unstable behavior and often leads to the formation of stop-and-go traffic jams for large enough number of string unstable drivers.

String instability can be mitigated by vehicle control using AVs or CAVs.
To demonstrate how vehicle control influences traffic, we simulate~(\ref{eq:dyn}, \ref{eq:HDM}, \ref{eq:ACC_general}) for a heterogeneous chain of HVs and AVs using ACC controller~(\ref{eq:ACC}) in Example~\ref{exmp:ACC}.
The simulation results shown by Fig.~\ref{fig:concept}(b) indicate that the AVs can successfully mitigate the onset of a traffic congestion: the tail vehicle brakes as much as the lead vehicle and not more (purple highlight).
We remark that congestions are often caused by the reaction time of human drivers: a delayed reaction forces drivers to overreact and brake or accelerate more.
As such, automated vehicles must be designed carefully to reduce their response delays, and their controllers must be tuned by considering the presence of delays.
There exist strategies for automated vehicles that specifically aim to eliminate the effect of delays~\cite{BekiarisLiberis2018, Molnar2018its, Qi2021, Wang2018a, Zhang2020}.
Still, even with well-designed automated vehicles it may require a relatively large penetration of AVs in the traffic flow to mitigate congestions if human drivers are string unstable (the penetration is 25\% in our example, i.e., every fourth vehicle is automated).
Connectivity and a CCC strategy could lead to further benefits: some level of connectivity may reduce the required penetration of automation to stabilize traffic~\cite{Avedisov2021}.
Yet, with vehicle control it is often hard to achieve stability for CAV penetrations around or below 10\%.

The benefits of connectivity are further exploited by traffic control strategies.
Simulation of~(\ref{eq:dyn}, \ref{eq:HDM}, \ref{eq:ATC_general}) with ATC controller~(\ref{eq:ATC}) in Example~\ref{exmp:ATC} is shown in Fig.~\ref{fig:concept}(c).
Here a single CAV responds to the vehicle ahead and to a single CHV ${N=10}$ vehicles behind.
With ATC strategy, the ego CAV is able to stabilize the traffic flow, and each vehicle including the CHV brakes less than the lead vehicle.
This is achieved with about 8\% penetration of automation and 17\% penetration of connectivity (one automated and two connected vehicles out of 12 vehicles).
The required penetration of automation is reduced (compared to when traffic is influenced by vehicle control) thanks to the extra connectivity --- which is a significant benefit since automation implies more cost and requirements than communication.
Moreover, the CHV has incentive to be connected to the CAV as it can slow down less.

The performance with and without connectivity is further explored in Fig.~\ref{fig:simulation}, where the influence of a single ACC vehicle on traffic is compared to traffic control by ATC.
These scenarios are illustrated in Fig.~\ref{fig:openloop}(b) and Fig.~\ref{fig:closedloop}(b); their only difference is that the vehicle $-N$ is connected in the ATC setup and the ego CAV responds to it.
It can be seen that ATC results in less speed reduction and smoother driving for the whole vehicle chain.
The price is that the headway of the CAV executing ATC fluctuates more than that for ACC.
However, it does not compromise safety, the vehicle is still far from collision.
In the meantime, smooth driving leads to less energy consumption, which is discussed next.

\subsection{Energy Efficiency}

We evaluate the energy consumption for each vehicle by the following measure~\cite{He2020}:
\begin{equation}
w_{n}(t) = \int_{t_0}^{t} v_{n}(\theta) g \big( \dot{v}_{n}(\theta) + p(v_{n}(\theta)) \big) {\rm d}\theta \,,
\label{eq:energy_consumption}
\end{equation}
that is, the energy over unit mass consumed during the time interval $[t_{0},t]$.
It is an integral of the power over unit mass obtained from the product of speed and commanded acceleration.
The latter involves the vehicle acceleration and the resistance terms represented by $p$ that vehicles need to overcome.
These can be modeled for example by
\begin{equation}
p(v) = a_{\rm r} + c_{\rm r} v^2 \,,
\label{eq:resistance}
\end{equation}
where $a_{\rm r}$ accounts for rolling resistance, while $c_{\rm r} v^2$ represents air drag.
Furthermore, function $g$ in~(\ref{eq:energy_consumption}) describes how the commanded acceleration is related to energy consumption.
For example, if we take
\begin{equation}
g(x) = \max\{ 0, x \} \,,
\label{eq:energy_ReLU}
\end{equation}
then we assume energy consumption during acceleration only and no energy consumption or regeneration during braking.

The energy measure~(\ref{eq:energy_consumption}, \ref{eq:resistance}, \ref{eq:energy_ReLU}) was evaluated for the simulations in Fig.~\ref{fig:simulation}; see the bottom of the plot.
The energy consumption stays constant during braking, increases linearly during constant speed cruising, and increases at a higher rate during acceleration.
Therefore, it is critical for the vehicles to drive smoothly and avoid decelerations after which they need to accelerate to recover the speed.

\begin{figure}[!t]
\begin{center}
\includegraphics[scale=.67]{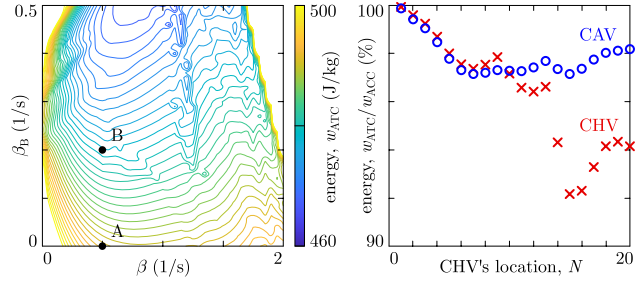}
\caption{
Total energy consumption of the ego CAV in the simulation scenario of Fig.~\ref{fig:simulation} with various $\beta$ and $\beta_{\rm B}$ control gains (left).
The energy consumption is reduced in traffic control (ATC, ${\beta_{\rm B} \neq 0}$) compared to vehicle control (ACC, ${\beta_{\rm B} = 0}$).
The relative energy benefits are shown for both the ego CAV and the CHV (right).}
\label{fig:energy}
\end{center}
\end{figure}

To highlight the difference between the total energy consumption of ACC and ATC, in Fig.~\ref{fig:energy} we plot the final value ${w_{0}(t_{\rm f})}$ of the energy consumed by the CAV over the simulation interval $[t_{0},t_{\rm f}]$.
Specifically, the simulations were repeated and the energy consumption was calculated for a grid of $\beta$ and $\beta_{\rm B}$ control gains representing various control designs.
The corresponding energy consumption levels are shown by the colored contours.
It can be seen that increasing $\beta_{\rm B}$ leads to a lower energy level.
One should, however, always keep safety in mind and choose $\beta_{\rm B}$ small enough to maintain a safe headway similarly to Fig.~\ref{fig:simulation}.
	
The gains corresponding to Fig.~\ref{fig:simulation} are indicated by point A for ACC (${\beta_{\rm B} = 0}$) and point B for ATC (${\beta_{\rm B} \neq 0}$) in Fig.~\ref{fig:energy}.
These two cases are compared on the right of Fig.~\ref{fig:energy} for various values of $N$.
It can be seen that ATC consistently saves around 2-3\% energy for the ego CAV compared to ACC if ${N \geq 5}$.
The energy consumption $w_{-N}(t_{\rm f})$ of the CHV is also indicated.
By deciding to stay connected, the CHV saves significant energy, around 6-8\% if ${N \geq 14}$.
The trend is that the farther the CHV is from the CAV (i.e., the larger $N$ is), the more energy it saves --- this is a consequence of a longer string unstable chain of HVs being more sensitive to the CAV's motion.

We remark that the energy contours in Fig.~\ref{fig:energy} are specific to the velocity profile of the lead vehicle and to the choice~(\ref{eq:energy_consumption}) of the energy measure.
Besides, extensive numerical simulations were needed to obtain these contours and to identify which controller parameters are energy efficient.
Finding energy-optimal parameters through theoretical analysis is a challenging task~\cite{He2020}.
On the other hand, energy consumption is associated with the smoothness of driving, thus it is related to the stability of the traffic flow.
String stable chains of vehicles attenuate speed fluctuations, drive smoother, and hence tend to consume less energy.
Stability analysis is more straightforward than finding energy optima.
Hence, below we analyze the dynamics and stability of traffic to drive the controller parameter selection.

\section{Dynamics of Traffic Flow}

In the remaining two sections, we study the dynamics and stability of vehicle and traffic control in detail.
The end result is manifested in so-called stability charts, which can be used to select the control gains of vehicle and traffic controllers such that string stability is achieved and traffic congestions are alleviated.
These two sections are algebraically more involved, thus readers who are less interested in the technical details can skip these and move to the Conclusions.

\subsection{Linearized Dynamics}

To analyze stability, first we consider traffic flows in equilibrium (often referred to as uniform flow) where all vehicles drive at the same constant speed.
The equilibrium speed is dictated by how fast the lead vehicle travels --- in practice it can be considered as the average speed of the lead vehicle --- and the following vehicles adjust their speed to the lead.
Afterwards, we consider deviations from this equilibrium to quantify how much speed perturbations amplify.
This will also be used to carry out stability analysis in the next section.
We conduct our analysis in the frequency domain by linearization and by constructing transfer functions that allow us to characterize the response of individual vehicles and the overall traffic behavior.

We start with formalizing the equilibrium solution as
\begin{equation}
v_{n}(t) \equiv v^* \,, \quad
s_{n}(t) = s_{n}^*(t) = v^* t + s_{n}(0) \,,
\label{eq:equilibrium}
\end{equation}
${n \in \mathcal{N}}$, which is associated with constant speed $v^*$ that is identical for each vehicle, and constant headway ${h_n^* = s_{n+1}^* - s_{n}^* - l_n}$ that may be different for each vehicle.
This equilibrium can be found by substituting~(\ref{eq:equilibrium}) into~(\ref{eq:dyn}, \ref{eq:HDM}) and~(\ref{eq:ACC_general}) for ACC or~(\ref{eq:ATC_general}) for ATC, and solving for $v^*$ and $h_{n}^*$.

We consider perturbations around the equilibrium in the form
\begin{equation}
v_{n}(t) = v^* + \tilde{v}_{n}(t) \,, \quad
s_{n}(t) = s_{n}^*(t) + \tilde{s}_{n}(t) \,,
\end{equation}
and collect position and speed perturbations into the state vector $\mathbf{x}$ given by
\begin{equation}
\mathbf{x}_{n}(t)
= \begin{bmatrix}
\tilde{s}_{n}(t) \\
\tilde{v}_{n}(t)
\end{bmatrix}
= \begin{bmatrix}
\tilde{v}_{n}(0) + \int_{0}^{t}\tilde{v}_{n}(\theta) {\rm d}\theta \\
\tilde{v}_{n}(t)
\end{bmatrix} ,
\end{equation}
from which the speed fluctuations can be recovered by
\begin{equation}
\tilde{v}_{n}(t) = \mathbf{c} \mathbf{x}_{n}(t), \quad
\mathbf{c}
= \begin{bmatrix}
0 & 1
\end{bmatrix}.
\label{eq:output}
\end{equation}

Assuming the accelerations of vehicles do not saturate, the linearized dynamics of HVs and CHVs can be derived from~(\ref{eq:dyn}, \ref{eq:HDM}) in the form
\begin{equation}
\dot{\mathbf{x}}_{n}(t) = \mathbf{a} \mathbf{x}_{n}(t) + \mathbf{a}_{\rm H} \mathbf{x}_{n}(t-\tau) + \mathbf{b}_{\rm H} \mathbf{x}_{n+1}(t-\tau) \,,
\label{eq:HV_linear}
\end{equation}
${n \in \{\mathcal{N}_{\rm HV}, \mathcal{N}_{\rm CHV}\}}$.
That is, it contains the delay-free state of vehicle $n$ and the delayed states of vehicles $n$ and $n+1$ with coefficient matrices
\begin{equation}
\mathbf{a}
= \begin{bmatrix}
0 & 1 \\
0 & 0
\end{bmatrix} , \quad
\mathbf{a}_{\rm H}
= \begin{bmatrix}
0 & 0 \\
- \left. \dfrac{\partial f_{\rm H}}{\partial h_{n}} \right|_{*} &
\left. \dfrac{\partial f_{\rm H}}{\partial v_{n}} \right|_{*}
- \left. \dfrac{\partial f_{\rm H}}{\partial \dot{h}_{n}} \right|_{*}
\end{bmatrix} , \quad
\mathbf{b}_{\rm H}
= \begin{bmatrix}
0 & 0 \\
\left. \dfrac{\partial f_{\rm H}}{\partial h_{n}} \right|_{*} &
\left. \dfrac{\partial f_{\rm H}}{\partial \dot{h}_{n}} \right|_{*}
\end{bmatrix} ,
\label{eq:HV_matrices_general}
\end{equation}
where star indicates that partial derivatives are evaluated at the equilibrium.
For AVs performing ACC the linearized dynamics become
\begin{equation}
\dot{\mathbf{x}}_{0}(t) = \mathbf{a} \mathbf{x}_{0}(t) + \mathbf{a}_{\rm F} \mathbf{x}_{0}(t-\sigma) + \mathbf{b}_{\rm F} \mathbf{x}_{1}(t-\sigma) \,,
\label{eq:ACC_linear}
\end{equation}
with
\begin{equation}
\mathbf{a}
= \begin{bmatrix}
0 & 1 \\
0 & 0
\end{bmatrix} , \quad
\mathbf{a}_{\rm F}
= \begin{bmatrix}
0 & 0 \\
\left. \dfrac{\partial f_{\text{ACC}}}{\partial s_{0}} \right|_{*} &
\left. \dfrac{\partial f_{\text{ACC}}}{\partial v_{0}} \right|_{*}
\end{bmatrix} , \quad
\mathbf{b}_{\rm F}
= \begin{bmatrix}
0 & 0 \\
\left. \dfrac{\partial f_{\text{ACC}}}{\partial s_{1}} \right|_{*} &
\left. \dfrac{\partial f_{\text{ACC}}}{\partial v_{1}} \right|_{*}
\end{bmatrix} .
\label{eq:ACC_matrices_general}
\end{equation}
whereas CAVs executing ATC are described by
\begin{equation}
\dot{\mathbf{x}}_{0}(t) = \mathbf{a} \mathbf{x}_{0}(t) + \mathbf{a}_{\rm FB} \mathbf{x}_{0}(t-\sigma) + \mathbf{b}_{\rm F} \mathbf{x}_{1}(t-\sigma) + \mathbf{b}_{\rm B} \mathbf{x}_{-N}(t-\sigma) \,,
\label{eq:ATC_linear}
\end{equation}
with
\begin{align}
\begin{split}
& \mathbf{a}
= \begin{bmatrix}
0 & 1 \\
0 & 0
\end{bmatrix} , \quad
\mathbf{a}_{\rm FB}
= \begin{bmatrix}
0 & 0 \\
\left. \dfrac{\partial f_{\text{ATC}}}{\partial s_{0}} \right|_{*} &
\left. \dfrac{\partial f_{\text{ATC}}}{\partial v_{0}} \right|_{*}
\end{bmatrix} , \\
& \mathbf{b}_{\rm F}
= \begin{bmatrix}
0 & 0 \\
\left. \dfrac{\partial f_{\text{ATC}}}{\partial s_{1}} \right|_{*} &
\left. \dfrac{\partial f_{\text{ATC}}}{\partial v_{1}} \right|_{*}
\end{bmatrix} , \quad
\mathbf{b}_{\rm B}
= \begin{bmatrix}
0 & 0 \\
\left. \dfrac{\partial f_{\text{ATC}}}{\partial s_{-N}} \right|_{*} &
\left. \dfrac{\partial f_{\text{ATC}}}{\partial v_{-N}} \right|_{*}
\end{bmatrix} .
\end{split}
\label{eq:ATC_matrices_general}
\end{align}

In summary, vehicle control influencing traffic is described by~(\ref{eq:HV_linear}, \ref{eq:ACC_linear}) at the linear level, while traffic control is characterized by~(\ref{eq:HV_linear}, \ref{eq:ATC_linear}).
For the ring configuration, one also has periodicity given by~(\ref{eq:periodicBC}) that yields
\begin{equation}
\mathbf{x}_{-N}(t) = \mathbf{x}_{1}(t) \,.
\label{eq:periodicBC_linear}
\end{equation}

\begin{example}
The coefficient matrices~(\ref{eq:HV_matrices_general}) describing human drivers have the following expressions for the case of the optimal velocity model~(\ref{eq:OVM}): 
\begin{equation}
\mathbf{a}
= \begin{bmatrix}
0 & 1 \\
0 & 0
\end{bmatrix}, \quad
\mathbf{a}_{\rm H}
= \begin{bmatrix}
0 & 0 \\
-\alpha_{\rm H} \kappa_{\rm H} & -(\alpha_{\rm H} + \beta_{\rm H})
\end{bmatrix}, \quad
\mathbf{b}_{\rm H}
= \begin{bmatrix}
0 & 0 \\
\alpha_{\rm H} \kappa_{\rm H} & \beta_{\rm H}
\end{bmatrix}.
\label{eq:HV_matrices}
\end{equation}
The ACC controller~(\ref{eq:ACC}) is associated with the coefficient matrices
\begin{equation}
\mathbf{a}
= \begin{bmatrix}
0 & 1 \\
0 & 0
\end{bmatrix}, \quad
\mathbf{a}_{\rm F}
= \begin{bmatrix}
0 & 0 \\
-\alpha \kappa & -(\alpha + \beta)
\end{bmatrix}, \quad
\mathbf{b}_{\rm F}
= \begin{bmatrix}
0 & 0 \\
\alpha \kappa & \beta
\end{bmatrix},
\label{eq:ACC_matrices}
\end{equation}
whereas the ATC controller~(\ref{eq:ATC}) corresponds to
\begin{equation}
\mathbf{a}
= \begin{bmatrix}
0 & 1 \\
0 & 0
\end{bmatrix}, \quad
\mathbf{a}_{\rm FB}
= \begin{bmatrix}
0 & 0 \\
-\alpha \kappa & -(\alpha + \beta + \beta_{\rm B})
\end{bmatrix}, \quad
\mathbf{b}_{\rm F}
= \begin{bmatrix}
0 & 0 \\
\alpha \kappa & \beta
\end{bmatrix}, \quad
\mathbf{b}_{\rm B}
= \begin{bmatrix}
0 & 0 \\
0 & \beta_{\rm B}
\end{bmatrix}.
\label{eq:ATC_matrices}
\end{equation}
\end{example}

\subsection{Transfer Functions}

We analyze the linearized dynamics in Laplace domain (assuming zero initial perturbations).
Our ultimate goal is to analyze the stability of the system, in particular, whether speed fluctuations amplify or decay as they propagate along the chain of vehicles.
This amplification property can be well represented by means of transfer functions describing the response of vehicles to the neighboring traffic.

First, we formulate the so-called {\em link transfer functions}~\cite{Zhang2016} associated with vehicle pairs.
For human drivers, the link transfer function $T_{\rm H}(s)$ relates the speed fluctuation of an HV to that of the vehicle ahead as
\begin{equation}
V_{n}(s) = T_{\rm H}(s) V_{n+1}(s) \,,
\end{equation}
where $T_{\rm H}(s)$ can be obtained from~(\ref{eq:output}, \ref{eq:HV_linear}) in the form
\begin{equation}
T_{\rm H}(s)
= \mathbf{c} \Big( s \mathbf{I} - \mathbf{a} - \mathbf{a}_{\rm H} {\rm e }^{-s \tau} \Big)^{-1} \mathbf{b}_{\rm H} {\rm e }^{-s \tau} 
\begin{bmatrix}
\frac{1}{s} \\ 1
\end{bmatrix} .
\label{eq:HV_TF_general}
\end{equation}
Similarly, in ACC where the automated vehicle responds to the vehicle immediately ahead only, the response can be characterized by a single link transfer function $T(s)$:
\begin{equation}
V_{0}(s) = T(s) V_{1}(s) \,,
\label{eq:ACC_response}
\end{equation}
whose expression can be derived from~(\ref{eq:ACC_linear}) as
\begin{equation}
T(s)
= \mathbf{c} \Big( s \mathbf{I} - \mathbf{a} - \mathbf{a}_{\rm F} {\rm e }^{-s \sigma} \Big)^{-1} \mathbf{b}_{\rm F} {\rm e }^{-s \sigma}
\begin{bmatrix}
\frac{1}{s} \\ 1
\end{bmatrix} .
\label{eq:ACC_TF_general}
\end{equation}

On the other hand, in ATC the ego CAV responds to multiple vehicles, associated with multiple link transfer functions.
For example, when responding to a single vehicle ahead and a single vehicle behind, we obtain
\begin{equation}
V_{0}(s) = T_{\rm F}(s) V_{1}(s) + T_{\rm B}(s) V_{-N}(s) \,,
\label{eq:ATC_response}
\end{equation}
where $T_{\rm F}(s)$ and $T_{\rm B}(s)$ are the link transfer functions characterizing the response forward and backward.
These are obtained from~(\ref{eq:ATC_linear}):
\begin{align}
\begin{split}
T_{\rm F}(s)
& = \mathbf{c} \Big( s \mathbf{I} - \mathbf{a} - \mathbf{a}_{\rm FB} {\rm e }^{-s \sigma} \Big)^{-1} \mathbf{b}_{\rm F} {\rm e }^{-s \sigma}
\begin{bmatrix}
\frac{1}{s} \\ 1
\end{bmatrix} , \\
T_{\rm B}(s)
& = \mathbf{c} \Big( s \mathbf{I} - \mathbf{a} - \mathbf{a}_{\rm FB} {\rm e }^{-s \sigma} \Big)^{-1} \mathbf{b}_{\rm B} {\rm e }^{-s \sigma}
\begin{bmatrix}
\frac{1}{s} \\ 1
\end{bmatrix} .
\end{split}
\label{eq:ATC_TF_general}
\end{align}
Notice that the denominators of $T_{\rm F}(s)$ and $T_{\rm B}(s)$ are the same, ${\rm D}(T_{\rm F}(s)) = {\rm D}(T_{\rm B}(s))$, since the same matrix is inverted.

Using the link transfer functions, we can analyze the overall response of traffic.
For a chain of $N$ human drivers, the overall response
\begin{equation}
V_{-N}(s) = \Gamma(s) V_{0}(s) \,.
\label{eq:HV_response}
\end{equation}
is characterized by $\Gamma(s)$ given by
\begin{equation}
\Gamma(s) = T_{\rm H}(s)^{N} \,.
\label{eq:HV_H2TTF}
\end{equation}
Note that this corresponds to identical human drivers; for the case of a heterogeneous chain of HVs it should be replaced with ${\Gamma(s) = \prod_{n = -N}^{-1} T_{{\rm H},n}(s)}$.

Introducing the ego vehicle (indexed $0$) and the vehicle ahead of it (lead vehicle~$1$) into the human-driven traffic, like in Figs.~\ref{fig:openloop}(b) and~\ref{fig:closedloop}(b), we can describe the overall response of the traffic flow from the head vehicle to the tail vehicle via the {\em head-to-tail transfer function}~\cite{Zhang2016} given by
\begin{equation}
V_{-N}(s) = G(s) V_{1}(s) \,.
\label{eq:response_overall}
\end{equation}
$G(s)$ depends on the human response $\Gamma(s)$ and the control strategy of the ego vehicle.
For vehicle control with ACC, we write~(\ref{eq:ACC_response}, \ref{eq:HV_response}) and obtain
\begin{equation}
G(s) = T(s) \Gamma(s) \,.
\label{eq:ACC_H2TTF}
\end{equation}
For the traffic control setup with ATC, we have~(\ref{eq:ATC_response}, \ref{eq:HV_response}) that lead to
\begin{equation}
G(s) = \frac{T_{\rm F}(s) \Gamma(s)}{1 - T_{\rm B}(s) \Gamma(s)} \,.
\label{eq:ATC_H2TTF}
\end{equation}
If traffic is considered on a ring road, we have the periodic boundary condition
\begin{equation}
V_{-N}(s) = V_{1}(s) \,,
\end{equation}
that leads to
\begin{equation}
G(s) = 1 \,,
\label{eq:ring_chareq}
\end{equation}
as the characteristic equation of the ring configuration.

\begin{remark}\textbf{-- Dynamics of rings and virtual rings.}
If vehicle control~(\ref{eq:ACC_H2TTF}) is executed on a ring road,~(\ref{eq:ring_chareq}) leads to
\begin{equation}
1 - T(s) \Gamma(s) = 0 \,.
\end{equation}
Here the left-hand side corresponds to the expression in the denominator of $G(s)$ in~(\ref{eq:ATC_H2TTF}).
Therefore, the dynamics of traffic control, that involves a virtual ring, is closely related to how vehicle control influences traffic when executed on a ring road.
As highlighted in the analysis below, the stability properties of the ring configuration give the backbone of the stability analysis of the virtual ring.
\end{remark}

\begin{example}\label{exmp:TF}
Substituting the coefficient matrices in~(\ref{eq:HV_matrices}) into the link transfer function~(\ref{eq:HV_TF_general}) describing human drivers leads to
\begin{equation}
T_{\rm H}(s)
= \frac{\beta_{\rm H} s + \alpha_{\rm H} \kappa_{\rm H}}{s^2 {\rm e}^{s \tau} + (\alpha_{\rm H} + \beta_{\rm H}) s + \alpha_{\rm H} \kappa_{\rm H}} \,.
\label{eq:HV_TF}
\end{equation}
Similarly, the link transfer function of ACC given by~(\ref{eq:ACC_matrices}) and~(\ref{eq:ACC_TF_general}) is
\begin{equation}
T(s)
= \frac{\beta s + \alpha \kappa}{s^2 {\rm e}^{s \sigma} + (\alpha + \beta) s + \alpha \kappa} \,,
\label{eq:ACC_TF}
\end{equation}
whereas for ATC~(\ref{eq:ATC_matrices}) and~(\ref{eq:ATC_TF_general}) give
\begin{align}
\begin{split}
T_{\rm F}(s)
& = \frac{\beta s + \alpha \kappa}{s^2 {\rm e}^{s \sigma} + (\alpha + \beta + \beta_{\rm B}) s + \alpha \kappa} \,, \\
T_{\rm B}(s)
& = \frac{\beta_{\rm B} s}{s^2 {\rm e}^{s \sigma} + (\alpha + \beta + \beta_{\rm B}) s + \alpha \kappa} \,.
\end{split}
\label{eq:ATC_TF}
\end{align}
Observe that we have ${T(s)=T_{\rm F}(s)/(1-T_{\rm B}(s))}$ that relates ACC and ATC.
\end{example}

\begin{remark}\textbf{-- Special cases.}
When the ego CAV does not respond to the traffic behind, ATC reduces to ACC.
Mathematically, this means setting ${T_{\rm B}(s)=0}$, ${T_{\rm F}(s) = T(s)}$, e.g. by taking ${\beta_{\rm B} = 0}$ in Example~\ref{exmp:TF}.
This reduces $G(s)$ in~(\ref{eq:ATC_H2TTF}) to $G(s)$ in~(\ref{eq:ACC_H2TTF}).
If no traffic is considered behind the ego CAV, then ${N=0}$ and ${\Gamma(s) = 1}$ due to~(\ref{eq:HV_H2TTF}).
In such a case, $G(s)$ in~(\ref{eq:ACC_H2TTF}) and~(\ref{eq:ATC_H2TTF}) both reduce to $T(s)$ describing the response of the ACC vehicle to the vehicle ahead only (since ${T(s)=T_{\rm F}(s)/(1-T_{\rm B}(s))}$ in our example).
Consequently, we can analyze the dynamics of ATC and then take these special cases to capture the dynamics of ACC followed by human-driven traffic or ACC on its own; recall Fig.~\ref{fig:interconnections}.
\end{remark}

\section{Stability of Traffic Flow}
\label{sec:stability}

Herein we formalize the stability conditions of traffic flows influenced by vehicle control and regulated by traffic control, while linking them to those of the ring configuration.

\subsection{Stability Conditions}

For a chain of vehicles traveling on a straight road, the system is non-autonomous, forced by the speed perturbations of the lead vehicle.
Thus, we consider the notions of {\em plant stability} and {\em string stability}~\cite{Feng2019, Ploeg2014a, Zhang2016} to analyze the behavior without speed perturbations and the response to the perturbations, respectively.
For the ring configuration, the system evolves autonomously, and we apply a single notion of stability; for more details on stability definitions see~\cite{Giammarino2019}.

String stability is particularly important in trarffic control.
It has various mathematical definitions; see a comprehensive summary in~\cite{Feng2019}.
Now we consider linear input-to-output string stability with respect to $\mathcal{L}_{2}$ norm.
This string stability notion is directly related to the magnitude ($\mathcal{H}_{\infty}$ norm) of the transfer function describing system, which makes it is easy to analyze and useful for control design.
Other notions are also widely-used, such as $\mathcal{L}_{p}$ string stability for any value of $p$~\cite{Ploeg2014a}.
$\mathcal{L}_{\infty}$ string stability is particularly relevant for traffic safety, since it characterizes the overshoot of the response to perturbations.
As this notion is related to the impulse response, it is more challenging to analyze it explicitly, and we rather consider $\mathcal{L}_{2}$ string stability.

\subsubsection{Plant Stability of a Vehicle Chain}
Plant stability means that each vehicle in the chain can approach a constant (equilibrium) speed in a stable manner.
If the vehicle chain is associated with head-to-tail transfer function $G(s)$, the plant stability condition is given by
\begin{equation}
{\rm D}(G(s_\ell)) = 0 \quad \Rightarrow \quad
\Re(s_\ell) < 0 \,, \quad
\forall \ell \in \mathbb{N} \,.
\label{eq:plant_stability}
\end{equation}
That is, the roots $s_{\ell}$ of the characteristic equation ${{\rm D}(G(s)) = 0}$ are located in the left-half of the complex plane, where ${\rm D}(G(s))$ is the denominator of the head-to-tail transfer function.
The system is at the plant stability boundary if either a real root ${s=0}$ is located on the imaginary axis:
\begin{equation}
{\rm D}({G(0)}) = 0 \,,
\label{eq:plant_stability_limit0}
\end{equation}
or a complex conjugate pair of roots ${s=\pm {\rm i} \Omega}$, ${\Omega > 0}$:
\begin{equation}
{\rm D}({G({\rm i} \Omega)}) = 0 \,.
\label{eq:plant_stability_limit}
\end{equation}

\subsubsection{String Stability of a Vehicle Chain}
String stability requires, on one hand, plant stability as a necessary condition, and on the other hand, that the speed perturbations of the head vehicle are not amplified by the vehicle chain.
Otherwise large speed fluctuations could arise at the end of the chain, ultimately leading to traffic jams and stop-and-go motion.
Specifically, we require that the speed fluctuations $V_{1}({\rm i} \omega)$ of the head vehicle with any given frequency ${\omega > 0}$ are smaller in amplitude than those $V_{-N}({\rm i} \omega)$ of the tail vehicle, which, based on~(\ref{eq:response_overall}), can be expressed as
\begin{equation}
\left| G({\rm i} \omega) \right| < 1 \,, \quad
\forall \omega > 0 \,.
\label{eq:string_stability}
\end{equation}
At the string stability boundary we have
\begin{equation}
\left| G({\rm i} \omega) \right| = 1 \quad \iff \quad
G({\rm i}\omega) = {\rm e}^{-{\rm i} K} \,,
\label{eq:string_stability_limit}
\end{equation}
for some ${\omega >0}$ and ${K \in [0, 2\pi)}$, i.e., the speed fluctuations of the head vehicle are "repeated" by the tail vehicle with a phase lag $K$ (also called as the wave number).

Furthermore, string stability can also be studied when ${\omega \to 0}$, shortly referred to as ${\omega = 0}$ string stability.
For algebraic convenience, we study this by introducing
\begin{equation}
P(\omega) = \frac{1}{\omega^2} \left( {\rm D} \left( \left| G({\rm i} \omega) \right|^2 \right) - {\rm N} \left( \left| G({\rm i} \omega) \right|^2 \right) \right) ,
\label{eq:string_stability_P}
\end{equation}
where ${\rm N}$ and ${\rm D}$ denote numerator and denominator, respectively.
With this definition,~(\ref{eq:string_stability}) can be rewritten as
\begin{equation}
P(\omega) > 0 \,, \quad
\forall \omega > 0 \,,
\end{equation}
and the ${\omega = 0}$ string stability boundary can be written simply as
\begin{equation}
P(0) = 0 \,.
\label{eq:string_stability_limit0}
\end{equation}

\subsubsection{Stability of the Ring Configuration}
Finally, the stability of the ring configuration is determined by the roots of the characteristic equation~(\ref{eq:ring_chareq}), and stability requires:
\begin{equation}
G(s_\ell) = 1 \quad \Rightarrow \quad
\Re(s_\ell) < 0 \,, \quad
\forall \ell \in \mathbb{N} \,,
\end{equation}
analogously to~(\ref{eq:plant_stability}).
Recall that $G(s)$ is the head-to-tail transfer function of the corresponding open chain configuration.
Since ${G(0) = 1}$ automatically holds, we consider the stability boundary where ${s = 0}$ is a double root of ${G(s) - 1}$, and where ${s=\pm {\rm i} \omega}$, $\omega>0$ satisfies the characteristic equation:
\begin{equation}
G({\rm i} \omega) = 1 \,.
\label{eq:ring_stability_limit}
\end{equation}

\subsection{Relationship of the Stability Conditions}
\label{sec:stability_connections}
Now let us take a deeper look into the relationship between the various vehicle configurations and their stability conditions.
These relationships are summarized at the bottom of Fig.~\ref{fig:interconnections} for the ${s={\rm i}\omega}$ stability boundaries.

For vehicle control with~(\ref{eq:ACC_H2TTF}), we have ${{\rm D}(G(s)) = {\rm D}(T(s)) {\rm D}(\Gamma(s))}$, hence the plant stability condition becomes
\begin{align}
\begin{split}
{\rm D}(T(s_\ell)) & = 0 \quad \Rightarrow \quad
\Re(s_\ell) < 0 \,, \quad
\forall \ell \in \mathbb{N} \,, \quad \text{and} \\
{\rm D}(\Gamma(s_\ell)) & = 0 \quad \Rightarrow \quad
\Re(s_\ell) < 0 \,, \quad
\forall \ell \in \mathbb{N} \,,
\end{split}
\label{eq:ACC_plant_stability}
\end{align}
that is, both the AV and HVs must be plant stable individually.
For string stability, we require
\begin{equation}
\left| T({\rm i} \omega) \right| \left| \Gamma({\rm i} \omega) \right| < 1 \,, \quad
\forall \omega > 0 \,,
\label{eq:ACC_string_stability}
\end{equation}
and at the stability boundary we have
\begin{equation}
T({\rm i} \omega) \Gamma({\rm i} \omega) = {\rm e}^{-{\rm i} K} \,,
\label{eq:ACC_string_stability_limit}
\end{equation}
which depend both on the response of the AV and the chain of HVs.
If the human-driven traffic is not considered, i.e., ${N = 0}$, ${\Gamma({\rm i} \omega) = 1}$, then we recover the string stability limit ${T({\rm i} \omega) = {\rm e}^{-{\rm i} K}}$ for a single ACC vehicle.

Now consider a nonzero (and potentially large) number $N$ of HVs.
This leads to one of the most interesting research questions, that several recent works have studied~\cite{Cui2017, Qin2020, Stern2018}: how many AVs are needed to mitigate traffic congestions?
If a human-driven vehicle chain behind an AV is long ($N$ is large), it results in a significant (exponential) increase of speed perturbations.
Thus, the AV has to make considerable effort to dampen the perturbation in order to prevent or mitigate a congestion.
Mathematically, the magnitude ${\left| T_{\rm H}({\rm i} \omega) \right|}$ of the individual human responses affects the overall response ${\left| \Gamma({\rm i} \omega) \right| = \left| T_{\rm H}({\rm i} \omega) \right|^N}$ more significantly for large $N$.
Therefore, it may be more challenging to stabilize a long chain of HVs with a single AV.

To highlight this, consider the limit ${N \to \infty}$.
If the human drivers are string stable, i.e., ${\left| T_{\rm H}({\rm i} \omega) \right| < 1}$, then ${\left| \Gamma({\rm i} \omega) \right| = 0}$ for all ${\omega > 0}$.
Unless ${\left| T({\rm i} \omega) \right| \to \infty}$, which can only happen at the plant stability boundary ${{\rm D}(T({\rm i} \omega)) = 0}$ of the AV,~(\ref{eq:ACC_string_stability}) is automatically satisfied.
Thus string stability is achieved for a long string stable chain of HVs and a plant stable AV.
However, when the human drivers are string unstable --- which is more likely the case in real traffic~\cite{Molnar2021trc} --- we have ${\left| T_{\rm H}({\rm i} \omega) \right| > 1}$ and ${\left| \Gamma({\rm i} \omega) \right| \to \infty}$ for some ${\omega > 0}$ and ${N \to \infty}$.
Thus, string stability in~(\ref{eq:ACC_string_stability}) cannot be achieved by any AV design (as long as ${T({\rm i} \omega) \neq 0}$, i.e., the AV responds to what is ahead).
This shows the fundamental limitation of influencing traffic by vehicle control: string stability cannot be achieved with arbitrarily small penetration of AVs (i.e., in the limit ${N \to \infty}$) if the human drivers are string unstable.

If vehicle control is executed on a ring, the stability limit~(\ref{eq:ring_stability_limit}) becomes
\begin{equation}
T({\rm i} \omega) \Gamma({\rm i} \omega) = 1 \,.
\label{eq:ACC_ring_chareq}
\end{equation}
Therefore, the stability boundary of the ring can be obtained as the ${K=0}$ special case of the string stability boundary~(\ref{eq:ACC_string_stability_limit}) of the open chain.
Physically, it means that perturbations propagate through the ring and arrive back to the same vehicle in the same phase due to the periodic boundary condition.
Another special case to mention is when all vehicles on the ring are identical, achieved by the substitution ${T(s) = T_{\rm H}(s)}$.
Then the characteristic equation~(\ref{eq:ACC_ring_chareq}) gives
\begin{equation}
T_{\rm H}({\rm i} \omega)^{N+1} = 1 \quad \iff \quad
T_{\rm H}({\rm i} \omega) = {\rm e}^{-{\rm i} \frac{2 k \pi}{N+1}} \,, \quad k \in \{0, 1, \ldots, N\} \,.
\label{eq:homogeneous_ring_chareq}
\end{equation}
When ${N \to \infty}$, the exponent ${K = 2 k \pi/(N+1)}$ may take any value on $[0, 2\pi)$ and we recover the string stability condition ${T_{\rm H}({\rm i} \omega) = {\rm e}^{-{\rm i} K}}$ for individual human drivers.
Thus, the stability of an infinitely long homogeneous ring is equivalent to the string stability of a homogeneous chain of vehicles.
Conversely, the substitution of ${K = 2 k \pi/(N+1)}$ into the string stability limit ${T_{\rm H}({\rm i} \omega) = {\rm e}^{-{\rm i} K}}$ of an individual human driver gives the stability limit~(\ref{eq:homogeneous_ring_chareq}) of the homogeneous ring.

\begin{figure}[!t]
\begin{center}
\includegraphics[scale=.67]{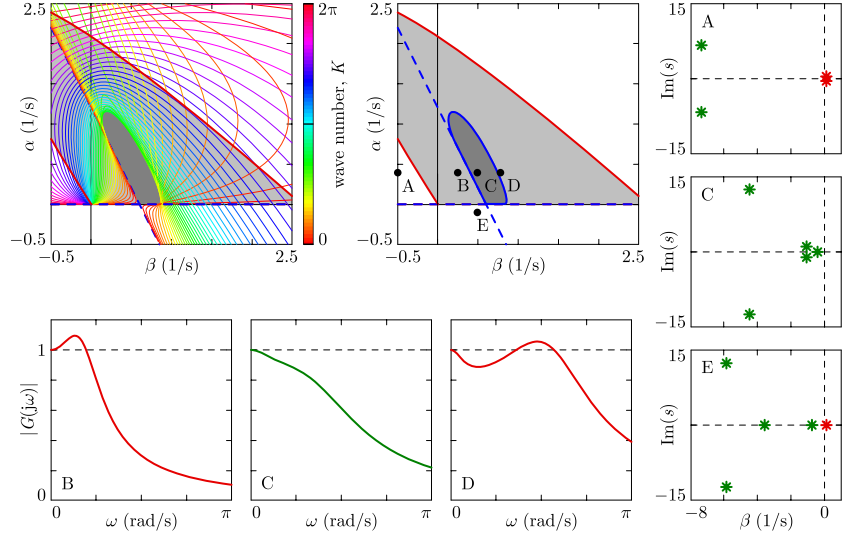}
\caption{
Stability chart of adaptive cruise control without considering its influence on the traffic behind (${N = 0}$).
Detailed (left) and simplified view (center).
The light gray region enclosed by red is plant stable, the dark gray region encircled by blue is string stable.
Characteristic roots (right) and frequency response (bottom) for specific control gains selected from the stability chart.
}
\label{fig:stabilitychart}
\end{center}
\end{figure}

Additionally, and most importantly, the traffic control setup as a virtual ring is related to both the open chain and ring configurations in terms of stability.
Consider formula~(\ref{eq:ATC_H2TTF}) of $G(s)$.
Since ${\rm D}(T_{\rm F}(s)) = {\rm D}(T_{\rm B}(s))$, we have 
\begin{equation}
{\rm D}(G(s))
= {\rm D}(T_{\rm B}(s) \Gamma(s)) \big( 1 - T_{\rm B}(s) \Gamma(s) \big) \,.
\end{equation}
Hence, and the plant stability condition~(\ref{eq:plant_stability}) leads to
\begin{align}
\begin{split}
{\rm D}(T_{\rm B}(s_\ell) \Gamma(s_\ell)) & = 0 \quad \Rightarrow \quad
\Re(s_\ell) < 0 \,, \quad
\forall \ell \in \mathbb{N} \,, \quad \text{and} \\
T_{\rm B}(s) \Gamma(s) & = 1 \quad \Rightarrow \quad
\Re(s_\ell) < 0 \,, \quad
\forall \ell \in \mathbb{N} \,.
\end{split}
\label{eq:ATC_plant_stability_decomposition}
\end{align}
That is, it requires both the plant stability of the associated vehicle control setup and the stability of the associated virtual ring configuration.

The string stability boundary of ATC can be written in the form
\begin{equation}
\frac{T_{\rm F}({\rm i} \omega) \Gamma({\rm i} \omega)}{1 - T_{\rm B}({\rm i} \omega) \Gamma({\rm i} \omega)} = {\rm e}^{-{\rm i} K} \,.
\end{equation}
This reduces to the string stability boundary of traffic flows influenced by vehicle control when there is no response to the traffic behind, i.e., ${T_{\rm B}({\rm i} \omega) = 0}$, ${T_{\rm F}({\rm i} \omega) = T({\rm i} \omega)}$, which is achieved by ${\beta_{\rm B} = 0}$ in our example.
In this sense, the string stability of ATC is related to that of ACC while the plant stability of ATC is related to the stability of its virtual ring.

\begin{figure}[!t]
\begin{center}
\includegraphics[scale=.67]{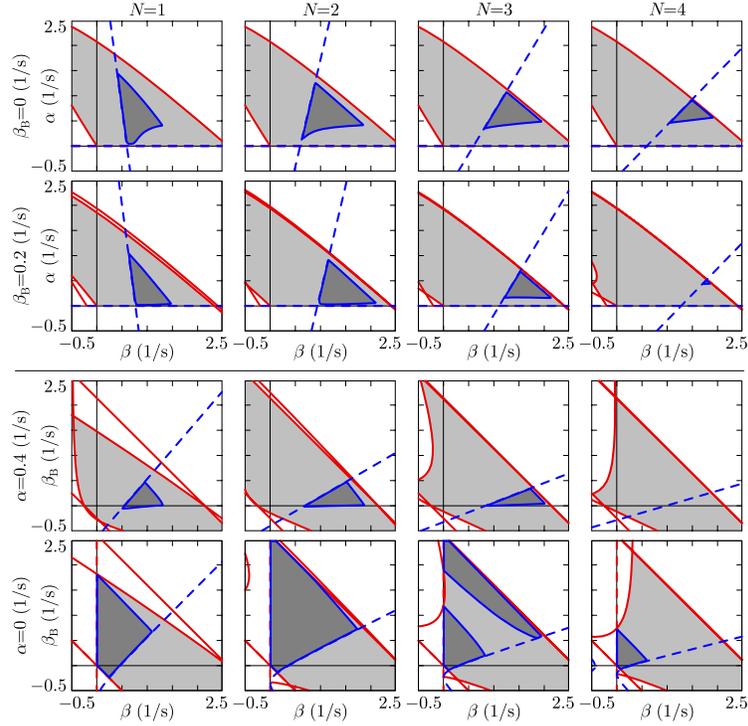}
\caption{
Stability charts of vehicle control influencing traffic (ACC, first row) and traffic control (ATC, second and third rows, and TC, fourth row).
The color scheme is the same as in Fig.~\ref{fig:stabilitychart}.
}
\label{fig:stabilitychart_sequence}
\end{center}
\end{figure}

\begin{example}
The stability analysis can be carried out for the specific choice~(\ref{eq:OVM}) of human driver model and selection~(\ref{eq:ACC}, \ref{eq:ATC}) of control laws.
The expressions~(\ref{eq:plant_stability_limit0}, \ref{eq:plant_stability_limit}) of the plant stability boundaries,~(\ref{eq:string_stability_limit}, \ref{eq:string_stability_limit0}) of the string stability boundaries and~(\ref{eq:ring_stability_limit}) of the stability boundaries for the ring can be expanded by using~(\ref{eq:HV_TF}, \ref{eq:ACC_TF}, \ref{eq:ATC_TF}).
Then, the stability boundaries can be expressed in terms of the controller parameters such as $\alpha$, $\beta$ or $\beta_{\rm B}$.
This allows one to select control gains such that they ensure stability.
The detailed formulas of this analysis are derived in the Appendix.

The resulting stability boundaries can be visualized in the space of control parameters, which we refer to as stability charts.
The left panel of Fig.~\ref{fig:stabilitychart} shows the stability chart in the $(\beta,\alpha)$ plane for ACC without considering the traffic behind, i.e., for ${N = 0}$.
This special case can also be found in~\cite{Zhang2016}.
Dashed red line indicates the ${s = 0}$ plant stability boundary, solid red line shows the ${s = \pm {\rm i} \Omega}$ plant stability boundary, and the plant stable region is light gray.
Furthermore, dashed blue lines show the ${\omega = 0}$ string stability boundaries (that overlap with the dashed red lines at ${\alpha = 0}$), and colorful curves denote the ${\omega > 0}$ string stability boundaries for various values of ${K \in [0, 2\pi)}$ indicated by color.
The envelope of these curves is presented in the center with a solid blue line, and the string stable region is dark gray.
The control gains shall be selected from this region to achieve stability.

Fig.~\ref{fig:stabilitychart} also illustrates the characteristics of traffic dynamics under different selections of the control gains $(\beta,\alpha)$.
The right panels show the characteristic roots associated with the plant stability condition~(\ref{eq:plant_stability}) for a plant unstable point with a complex pair of unstable roots (A), a plant stable point (C), and a plant unstable point with a real unstable root (E).
The bottom panels depict the magnitude of the head-to-tail transfer function related to the string stability condition~(\ref{eq:string_stability}) for a point with ${\omega = 0}$ string instability (B), a string stable point (C) and a point with ${\omega > 0}$ string instability (D).
Out of these points, only point C has desired behavior.
This illustrates how stability charts may drive the selection of controller parameters, so that the end result is a stable and smooth traffic flow.

The stability charts also given another aspect of comparing vehicle and traffic control strategies.
The top row of Fig.~\ref{fig:stabilitychart_sequence} show the stability charts in the $(\beta,\alpha)$ plane for ACC followed by various numbers of human drivers (${N = 1, 2, 3, 4}$).
Since the human driver parameters are selected to be string unstable, the string stable region shrinks as the number of vehicles increases.
Furthermore, it also shifts towards large gains that may require large control input and acceleration from the ACC vehicle.
In comparison, the string stable region of ATC in the second row of Fig.~\ref{fig:stabilitychart_sequence} shrinks faster, but is located at lower ${\alpha}$ gains associated with smaller acceleration and better passenger comfort.

For a fixed, reasonably small gain ${\alpha}$, the stability charts of ATC are plotted in the $(\beta,\beta_{\rm B})$ plane in the third row of Fig.~\ref{fig:stabilitychart_sequence}.
Note that the points along ${\beta_{\rm B} = 0}$ represent ACC as a special case, and they are located close to or outside the string stable region.
Hence for such realistic choice of ${\alpha}$, traffic control with ATC (${\beta_{\rm B} \neq 0}$) may achieve string stability while vehicle control with ACC (${\beta_{\rm B} = 0}$) may not.
For a large number of string unstable human drivers, the string stable region disappears (see ${N = 4}$).
However, ATC may still be beneficial by reducing instability, improving the smoothness of driving behavior, and reducing energy consumption; as these were highlighted in Figs.~\ref{fig:concept} and~\ref{fig:energy}.

Finally, the last row of Fig.~\ref{fig:stabilitychart_sequence} shows the stability charts of TC (as the ${\alpha=0}$ special case of ATC).
Although the string stable region shrinks as the chain of HVs gets longer ($N$ gets larger), it is significantly larger than that of ATC.
In fact, the string stable region does not vanish even for chains as long as ${N=10}$ HVs.
This indicates that CAVs can control traffic significantly more efficiently in situations where they do not have to respond to the position of the vehicle ahead.
\end{example}

\section{Conclusions}

We have discussed vehicle control strategies influencing traffic and traffic control strategies regulating traffic by means of injecting connected automated vehicles (CAVs) into the traffic flow.
We have shown that vehicle-to-everything (V2X) connectivity allows a CAV to receive information about the traffic behind it, respond to it, and mitigate traffic congestions, while also responding to vehicles ahead.
The response to the traffic behind creates a feedback loop for traffic control, in a similar structure to a ring road, termed virtual ring.
This strategy achieves smooth driving and energy savings for both the CAV, and the vehicles behind it that decide to stay connected to help traffic control.
We have also analyzed the stability of the traffic flow and derived stability charts that guide controller tuning.

To realize the proposed traffic control strategies on real highways, it is essential for the CAV to access information about the traffic behind it.
Thus, sufficient penetration of connectivity is required so that the CAV can sync with connected vehicles behind it in its communication range, that may be limited to a few hundred meters.
Furthermore, the controllers proposed in this paper were discussed for a single-lane scenario --- for multiple lanes they should be extended by considering the cross-lane dynamics, and lane information should be acquired and incorporated into the control laws.
Finally, we emphasize that connected vehicles behind the CAV do not necessarily have to be automated.
If they possess sufficient level of automation, however, they can coordinate with the CAV for more efficient traffic control.
Our future work will explore coordination of multiple CAVs and multi-lane dynamics in traffic control.

\begin{acknowledgement}
This work was supported by Ford Motor Co. and the University of Michigan's Center for Connected and Automated Transportation through the US DOT grant 69A3551747105.
\end{acknowledgement}

\section*{Appendix}
\addcontentsline{toc}{section}{Appendix}

The parameters used throughout the paper are given in Table~\ref{tab:parameters}.
The detailed formulas of the stability boundaries plotted in Sec.~\ref{sec:stability} are given below.

\begin{table}
\caption{Parameters of the numerical case studies.}
\label{tab:parameters}
\begin{center}
\begin{tabular}{c c c c}
\hline
Field & Description & Parameter & Value \\
\hline
\multirow{3}{*}{vehicle properties}
& vehicle length & $l$ & 5 m \\
& braking limit & $a_{\rm min}$ & 7 m/s$^2$ \\
& acceleration limit & $a_{\rm max}$ & 3 m/s$^2$ \\
\hline
\multirow{5}{*}{human-driven vehicle}
& time delay & $\tau$ & 0.8 s$^{-1}$ \\
& gain for headway response & $\alpha_{\rm H}$ & 0.1 s$^{-1}$ \\
& gain for speed response & $\beta_{\rm H}$ & 0.6 s$^{-1}$ \\
& range policy & $F_{\rm H}(h)$ & ${v_{\rm max} \left( 1 - \left( \frac{h_{\rm go} - h}{h_{\rm go} - h_{\rm st}} \right)^2 \right)}$ \\
\hline
\multirow{5}{*}{automated vehicle}
& time delay & $\sigma$ & 0.6 s$^{-1}$ \\
& gain for headway response & $\alpha$ & 0.4 s$^{-1}$ \\
& gain for speed response & $\beta$ & 0.5 s$^{-1}$ \\
& gain for traffic control & $\beta_{\rm B}$ & 0.2 s$^{-1}$ \\
& range policy & $F(h)$ & ${v_{\rm max} \frac{h - h_{\rm st}}{h_{\rm go} - h_{\rm st}}}$ \\
\hline
\multirow{3}{*}{range policy}
& standstill headway & $h_{\rm st}$ & 5 m \\
& free flow headway & $h_{\rm go}$ & 55 m \\
& speed limit & $v_{\rm max}$ & 30 m/s \\
\hline
\multirow{6}{*}{simulations}
& time interval & ${[t_{0},t_{\rm f}]}$ & [0,60] s \\
& time step & $\Delta t$ & 0.01 s \\
& lead vehicle braking & $a_{1}(t)$ & -1 m/s$^2$, ${t \in [0,10]}$ s \\
& lead vehicle acceleration & $a_{1}(t)$ & 0.5 m/s$^2$, ${t \in [10,30]}$ s \\
& \multirow{2}{*}{initial conditions} & $s_{n}(t)$, $v_{n}(t)$ & \multirow{2}{*}{constant $s_{n}^*$, $v_{n}^*$} \\ & & $t \in [-\tau_{n},0]$ & \\
\hline
\multirow{3}{*}{energy consumption}
& constant resistance coefficient & $a_{\rm r}$ & 0.0981 m/s$^2$ \\
& quadratic resistance coefficient & $c_{\rm r}$ & 0.0003 m$^{-1}$ \\
& resolution of contour plots & ${\Delta \beta \times \Delta \beta_{\rm B}}$ & ${0.05 \times 0.01}$ s$^{-1}$ \\
\hline
\multirow{4}{*}{stability charts}
& range policy gradient for HVs & $\kappa_{\rm H}$ & 0.7 s$^{-1}$ \\
& range policy gradient for AVs & $\kappa$ & 0.6 s$^{-1}$ \\
& frequency range for plant stability & ${[\Omega_{\rm min},\Omega_{\rm max}]}$ & $[0,2\pi]$ rad/s \\
& frequency range for string stability & ${[\omega_{\rm min},\omega_{\rm max}]}$ & $[0,2\pi]$ rad/s \\
\hline
\end{tabular}
\end{center}
\end{table}

\subsection*{Stability of ATC}

As highlighted in Fig.~\ref{fig:interconnections}, the car-following configurations in our examples can all be considered as the special cases of traffic control with ATC.
Therefore, we derive the stability charts of ATC and then obtain those of other scenarios as special cases.
Throughout the appendix we use~(\ref{eq:HV_TF}, \ref{eq:ACC_TF}, \ref{eq:ATC_TF}) as the expressions of the link transfer functions $T_{\rm H}(s)$, $T(s)$, $T_{\rm F}(s)$ and $T_{\rm B}(s)$. 

First, consider the plant stability of ATC.
Based on~(\ref{eq:ATC_plant_stability_decomposition}), it decomposes into the plant stability of the associated vehicle control setup and virtual ring.
Consider the first line of~(\ref{eq:ATC_plant_stability_decomposition}) and assume a plant stable chain of HVs described by $\Gamma(s)$ (the conditions of this are given at the end of the Appendix).
Then we can take ${{\rm D}(T_{\rm B}(s)) = 0}$, which leads to the characteristic equation
\begin{equation}
s^2 {\rm e}^{s \sigma} + (\alpha + \beta + \beta_{\rm B}) s + \alpha \kappa = 0 \,.
\end{equation}
The case ${s = 0}$ gives the plant stability boundary
\begin{equation}
\alpha = 0 \,,
\label{eq:ATC_plant_1}
\end{equation}
whereas substitution of ${s = \pm {\rm i} \Omega}$, separation into real and imaginary parts and solution for $\alpha$ and $\beta$ lead to the plant stability boundary
\begin{align}
\begin{split}
\alpha & = \frac{\Omega^2 \cos(\Omega \sigma)}{\kappa} \,, \\
\beta & = \Omega \sin(\Omega \sigma) - \alpha - \beta_{\rm B}\,.
\end{split}
\label{eq:ATC_plant_2}
\end{align}
This can also be rearranged to
\begin{equation}
\beta_{\rm B} = \Omega^*(\alpha) \sin \big( \Omega^*(\alpha) \sigma \big) - \alpha - \beta \,,
\end{equation}
where $\Omega^*(\alpha)$ is the solution of ${\alpha = \Omega^2 \cos(\Omega \sigma)/\kappa}$ for $\Omega$.

Similarly, the second line of~(\ref{eq:ATC_plant_stability_decomposition}) leads to the characteristic equation
\begin{equation}
s^2 {\rm e}^{s \sigma} + (\alpha + \beta + \beta_{\rm B}) s + \alpha \kappa - \beta_{\rm B} s \Gamma(s) = 0 \,.
\end{equation}
For ${s = 0}$, this yields the plant stability boundary
\begin{equation}
\alpha = 0 \,,
\label{eq:ATC_plant_3}
\end{equation}
while for ${s = \pm {\rm i} \Omega}$ it implies
\begin{align}
\begin{split}
\alpha & = \frac{\Omega^2 \cos(\Omega \sigma) - \Omega \Gamma_{\rm I} \beta_{\rm B}}{\kappa} \,, \\
\beta & = \Omega \sin(\Omega \sigma) - \alpha - (1 - \Gamma_{\rm R}) \beta_{\rm B} \,,
\end{split}
\label{eq:ATC_plant_4}
\end{align}
or, equivalently,
\begin{align}
\begin{split}
\beta_{\rm B} & = \frac{\Omega^2 \cos(\Omega \sigma) - \alpha \kappa}{\Omega \Gamma_{\rm I}} \,, \\
\beta & = \Omega \sin(\Omega \sigma) - \alpha - (1 - \Gamma_{\rm R}) \beta_{\rm B} \,,
\end{split}
\end{align}
where ${\Gamma_{\rm R} = \Re(\Gamma({\rm i} \Omega))}$ and ${\Gamma_{\rm I} = \Im(\Gamma({\rm i} \Omega))}$.
These equations were used to plot the plant stability boundaries in the $(\beta,\alpha)$ and $(\beta,\beta_{\rm B})$ planes in Figs.~\ref{fig:stabilitychart} and~\ref{fig:stabilitychart_sequence}.

String stability can be analyzed by the help of the head-to-tail transfer function~(\ref{eq:ATC_H2TTF}), which now reads
\begin{equation}
G({\rm i}\omega) = \frac{(\beta {\rm i} \omega + \alpha \kappa) (\Gamma_{\rm R} + {\rm i} \Gamma_{\rm I})}{-\omega^2 {\rm e}^{{\rm i} \omega \sigma} + (\alpha + \beta + \beta_{\rm B}) {\rm i} \omega + \alpha \kappa - \beta_{\rm B} {\rm i} \omega (\Gamma_{\rm R} + {\rm i} \Gamma_{\rm I})} \,,
\end{equation}
for ${s = {\rm i} \omega}$, where ${\Gamma_{\rm R} = \Re(\Gamma({\rm i} \omega))}$ and ${\Gamma_{\rm I} = \Im(\Gamma({\rm i} \omega))}$.
Direct substitution into~(\ref{eq:string_stability_P}) gives the following expression for $P(\omega)$:
\begin{multline}
P(\omega) = \omega^2 - 2 \alpha \kappa \cos(\omega \sigma) - 2 \beta_{\rm B} \omega \Gamma_{\rm I} \cos(\omega \sigma) + \alpha^2 \kappa^2 \frac{1 - \Gamma_{\rm R}^2 - \Gamma_{\rm I}^2}{\omega^2} + 2 \alpha \kappa \beta_{\rm B} \frac{\Gamma_{\rm I}}{\omega} + \beta_{\rm B}^2 \Gamma_{\rm I}^2 \\
- 2 \big( \alpha + \beta + (1 - \Gamma_{\rm R}) \beta_{\rm B} \big) \omega \sin(\omega \sigma) + \big( \alpha + \beta + (1 - \Gamma_{\rm R}) \beta_{\rm B} \big)^2 - \beta^2(\Gamma_{\rm R}^2 + \Gamma_{\rm I}^2) \,.
\end{multline}

We can construct the ${\omega=0}$ string stability boundaries via~(\ref{eq:string_stability_limit0}).
Note that the following hold for ${\omega \to 0}$:
\begin{align}
\begin{split}
\lim_{\omega \to 0} \frac{\Gamma_{\rm I}}{\omega} & = - \frac{N}{\kappa_{\rm H}} \,, \\
\lim_{\omega \to 0} \frac{\Gamma_{\rm R} - 1}{\omega^2} & = N \frac{2 \kappa_{\rm H} - 2 \beta_{\rm H} - (N+1) \alpha_{\rm H}}{2 \alpha_{\rm H} \kappa_{\rm H}^2} \,.
\end{split}
\end{align}
This can be shown by substituting~(\ref{eq:HV_TF}) into~(\ref{eq:HV_H2TTF}), taking ${s = {\rm i} \omega}$, and calculating the limits above by applying L'Hospital's rule (once for $\Gamma_{\rm I}/\omega$ and twice for ${(\Gamma_{\rm R} - 1)/\omega^2}$).
At the limit ${\omega \to 0}$ we get
\begin{equation}
P(0) = \alpha \bigg( \alpha + 2 \beta - 2 \kappa + N \frac{\alpha \kappa^2}{\alpha_{\rm H} \kappa_{\rm H}^2} (\alpha_{\rm H} + 2 \beta_{\rm H} - 2 \kappa_{\rm H}) - 2 N \frac{\kappa}{\kappa_{\rm H}} \beta_{\rm B} \bigg) \,.
\end{equation}
Taking ${P(0)=0}$ finally gives the ${\omega = 0}$ string stability boundaries in the form
\begin{align}
\begin{split}
\alpha & = 0 \,, \\
\alpha & = \frac{2 \Big( \kappa - \beta + N \frac{\kappa}{\kappa_{\rm H}} \beta_{\rm B} \Big)}{1 + \frac{N \kappa^2}{\alpha_{\rm H} \kappa_{\rm H}^2} (\alpha_{\rm H} + 2 \beta_{\rm H} - 2 \kappa_{\rm H})} \,,
\end{split}
\end{align}
which can also be expressed as
\begin{equation}
\beta_{\rm B} = \frac{\kappa_{\rm H}}{2 N \kappa} \bigg( \alpha + 2 \beta - 2 \kappa + N \frac{\alpha \kappa^2}{\alpha_{\rm H} \kappa_{\rm H}^2} (\alpha_{\rm H} + 2 \beta_{\rm H} - 2 \kappa_{\rm H}) \bigg) \,.
\end{equation}

The ${\omega > 0}$ string stability boundaries can be found by substitution of~(\ref{eq:ATC_H2TTF}) into~(\ref{eq:string_stability_limit}).
Then, separation into real and imaginary parts leads to a linear system of equations for $\alpha$ and $\beta$, that ultimately gives
\begin{align}
\begin{split}
\alpha & = \frac{\omega^2 \big( \Gamma_{\rm I} \sin(\omega \sigma - K) + \Gamma_{\rm R} \cos(\omega \sigma - K) - \cos(\omega \sigma) \big)}
{\omega (\Gamma_{\rm R} \sin K + \Gamma_{\rm I} \cos K) - \kappa ( 1 + \Gamma_{\rm R}^2 + \Gamma_{\rm I}^2 - 2 \Gamma_{\rm R} \cos K + 2 \Gamma_{\rm I} \sin K )} \\
& \quad + \frac{\beta_{\rm B} \omega \big( (\Gamma_{\rm R}^2 + \Gamma_{\rm I}^2) \sin K - \Gamma_{\rm R} \sin K + \Gamma_{\rm I} (1 - \cos K) \big)}
{\omega (\Gamma_{\rm R} \sin K + \Gamma_{\rm I} \cos K) - \kappa ( 1 + \Gamma_{\rm R}^2 + \Gamma_{\rm I}^2 - 2 \Gamma_{\rm R} \cos K + 2 \Gamma_{\rm I} \sin K )} \,, \\
\beta & = \frac{\omega^2 \cos(\omega \sigma) - \kappa \omega \big( \Gamma_{\rm I} \cos(\omega \sigma - K) - \Gamma_{\rm R} \sin(\omega \sigma - K) + \sin(\omega \sigma) \big)}
{\omega (\Gamma_{\rm R} \sin K + \Gamma_{\rm I} \cos K) - \kappa ( 1 + \Gamma_{\rm R}^2 + \Gamma_{\rm I}^2 - 2 \Gamma_{\rm R} \cos K + 2 \Gamma_{\rm I} \sin K )} \\
& \quad + \frac{\beta_{\rm B} \big( -\omega \Gamma_{\rm I} + \kappa \big( 1 + (\Gamma_{\rm R}^2 + \Gamma_{\rm I}^2) \cos K - \Gamma_{\rm R} (1 + \cos K) + \Gamma_{\rm I} \sin K \big) \big)}
{\omega (\Gamma_{\rm R} \sin K + \Gamma_{\rm I} \cos K) - \kappa ( 1 + \Gamma_{\rm R}^2 + \Gamma_{\rm I}^2 - 2 \Gamma_{\rm R} \cos K + 2 \Gamma_{\rm I} \sin K )} \,.
\end{split}
\label{eq:ATC_string}
\end{align}
This can be re-written also in the form
\begin{align}
\begin{split}
\beta_{\rm B} & = \frac{\omega^2 \big( - \Gamma_{\rm I} \sin(\omega \sigma - K) - \Gamma_{\rm R} \cos(\omega \sigma - K) + \cos(\omega \sigma) ) \big)}
{\omega \big( \Gamma_{\rm I} (1 - \cos K) - \Gamma_{\rm R} \sin K + (\Gamma_{\rm R}^2 + \Gamma_{\rm I}^2) \sin K \big)} \\
& \quad + \frac{ \alpha \omega (\Gamma_{\rm R} \sin K + \Gamma_{\rm I} \cos K) - \alpha \kappa \big( 1 + \Gamma_{\rm R}^2 + \Gamma_{\rm I}^2 - 2 \Gamma_{\rm R} \cos K + 2 \Gamma_{\rm I} \sin K \big)}
{\omega \big( \Gamma_{\rm I} (1 - \cos K) - \Gamma_{\rm R} \sin K + (\Gamma_{\rm R}^2 + \Gamma_{\rm I}^2) \sin K \big)} \,, \\
\beta & = \frac{\omega^2 \big( \Gamma_{\rm I} \sin(\omega \sigma) - (1 - \Gamma_{\rm R}) \cos(\omega \sigma) \big)}
{\omega \big( \Gamma_{\rm I} (1 - \cos K) - \Gamma_{\rm R} \sin K + (\Gamma_{\rm R}^2 + \Gamma_{\rm I}^2) \sin K \big)} \\
& \quad + \frac{-\alpha \omega \Gamma_{\rm I} + \alpha \kappa \big( 1 + (\Gamma_{\rm R}^2 + \Gamma_{\rm I}^2) \cos K - \Gamma_{\rm R} (1 + \cos K) + \Gamma_{\rm I} \sin K \big)}
{\omega \big( \Gamma_{\rm I} (1 - \cos K) - \Gamma_{\rm R} \sin K + (\Gamma_{\rm R}^2 + \Gamma_{\rm I}^2) \sin K \big)} \,.
\end{split}
\end{align}
These formulas allowed us to plot the string stability boundaries in the $(\beta,\alpha)$ and $(\beta,\beta_{\rm B})$ planes in Figs.~\ref{fig:stabilitychart} and~\ref{fig:stabilitychart_sequence}.

\subsection*{Stability of ACC}

Now we consider the stability of traffic flows influenced by vehicle control via ACC.

\subsubsection*{ACC Followed by Human-Driven Traffic}
Since ACC is a special case of ATC, all stability boundaries can be obtained by substituting ${\beta_{\rm B} = 0}$.
The plant stability boundaries~(\ref{eq:ATC_plant_1})-(\ref{eq:ATC_plant_4}) reduce to
\begin{equation}
\alpha = 0 \,,
\label{eq:ACC_plant0}
\end{equation}
and
\begin{align}
\begin{split}
\alpha & = \frac{\Omega^2 \cos(\Omega \sigma)}{\kappa} \,, \\
\beta & = \Omega \sin(\Omega \sigma) - \alpha \,,
\label{eq:ACC_plant}
\end{split}
\end{align}
whereas the string stability boundaries become
\begin{align}
\begin{split}
\alpha & = 0 \,, \\
\alpha & = \frac{2(\kappa - \beta)}{1 + \frac{N \kappa^2}{\alpha_{\rm H} \kappa_{\rm H}^2} (\alpha_{\rm H} + 2 \beta_{\rm H} - 2 \kappa_{\rm H})} \,,
\end{split}
\label{eq:ACC_string0}
\end{align}
and
\begin{align}
\begin{split}
\alpha & = \frac{\omega^2 \big( \Gamma_{\rm I} \sin(\omega \sigma - K) + \Gamma_{\rm R} \cos(\omega \sigma - K) - \cos(\omega \sigma) \big)}{\omega (\Gamma_{\rm R} \sin K + \Gamma_{\rm I} \cos K) - \kappa ( 1 + \Gamma_{\rm R}^2 + \Gamma_{\rm I}^2 - 2 \Gamma_{\rm R} \cos K + 2 \Gamma_{\rm I} \sin K )} \,, \\
\beta & = \frac{\omega^2 \cos(\omega \sigma) - \kappa \omega \big( \Gamma_{\rm I} \cos(\omega \sigma - K) - \Gamma_{\rm R} \sin(\omega \sigma - K) + \sin(\omega \sigma) \big)}{\omega (\Gamma_{\rm R} \sin K + \Gamma_{\rm I} \cos K) - \kappa ( 1 + \Gamma_{\rm R}^2 + \Gamma_{\rm I}^2 - 2 \Gamma_{\rm R} \cos K + 2 \Gamma_{\rm I} \sin K )} \,.
\end{split}
\label{eq:ACC_string}
\end{align}

\subsubsection*{ACC Followed by Human-Driven Traffic on a Ring}

When the ACC scenario with subsequent human-driven vehicles is driven on a ring road, the characteristic equation becomes~(\ref{eq:ring_chareq}).
As pointed out in Sec.~\ref{sec:stability_connections}, the stability boundary is analogous to the plant stability of the virtual ring in ATC.
Thus, for ${s=0}$ we obtain
\begin{equation}
\alpha = 0 \,,
\end{equation}
cf.~(\ref{eq:ATC_plant_1}, \ref{eq:ATC_plant_3}).
For ${s={\rm i} \omega}$, the stability boundary can be obtained as the ${K=0}$ special case of the ${\omega>0}$ string stability boundary of the associated straight-road configuration.
Thus, substituting ${K=0}$ into~(\ref{eq:ACC_string}) leads to
\begin{align}
\begin{split}
\alpha & = \frac{\omega^2 \big( \Gamma_{\rm I} \sin(\omega \sigma) - (1 - \Gamma_{\rm R}) \cos(\omega \sigma) \big)}{\omega \Gamma_{\rm I} - \kappa \big( (1 - \Gamma_{\rm R})^2 + \Gamma_{\rm I}^2 \big)} \,, \\
\beta & = \frac{\omega^2 \cos(\omega \sigma) - \kappa \omega \big( \Gamma_{\rm I} \cos(\omega \sigma) + (1 - \Gamma_{\rm R}) \sin(\omega \sigma) \big)}{\omega \Gamma_{\rm I} - \kappa \big( (1 - \Gamma_{\rm R})^2 + \Gamma_{\rm I}^2 \big)} \,.
\end{split}
\end{align}
Notice that this gives back the plant stability boundaries~(\ref{eq:ACC_plant}) if we substitute ${\Gamma({\rm i} \omega) = 0}$, i.e., ${\Gamma_{\rm R} = 0}$, ${\Gamma_{\rm I} = 0}$ (which is the special case of ${N \to \infty}$ string stable human drivers with ${\left| T_{\rm H}({\rm i} \omega) \right| < 1}$, see Sec.~\ref{sec:stability_connections}).

\subsubsection*{ACC without Considering the Traffic Behind}

The influence of ACC controllers on the traffic flow behind them is often neglected during cruise control design.
This can be considered as the ${N = 0}$ special case of ACC followed by human-driven traffic discussed above.
For ${N = 0}$, we have ${\Gamma(s) = 1}$ that implies
${\Gamma_{\rm R} = 1}$, ${\Gamma_{\rm I} = 0}$.
The plant stability conditions still yield~(\ref{eq:ACC_plant0}, \ref{eq:ACC_plant}), since we have already considered a plant stable human-driven chain:
\begin{equation}
\alpha = 0 \,,
\end{equation}
and
\begin{align}
\begin{split}
\alpha & = \frac{\Omega^2 \cos(\Omega \sigma)}{\kappa} \,, \\
\beta & = \Omega \sin(\Omega \sigma) - \alpha \,.
\end{split}
\end{align}
On the other hand, substitution of ${N = 0}$ into~(\ref{eq:ACC_string0}) leads to the ${\omega=0}$ string stability boundaries
\begin{align}
\begin{split}
\alpha & = 0 \,, \\
\alpha & = 2 (\kappa - \beta) \,,
\end{split}
\end{align}
while substitution of ${\Gamma_{\rm R} = 1}$ and ${\Gamma_{\rm I} = 0}$ into~(\ref{eq:ACC_string}) results in the ${\omega > 0}$ string stability boundaries
\begin{align}
\begin{split}
\alpha & = \frac{\omega^2 \big( \cos(\omega \sigma - K) - \cos(\omega \sigma) \big)}
{\omega \sin K - 2 \kappa (1- \cos K)} \,, \\
\beta & = \frac{\omega^2 \cos(\omega \sigma) + \kappa \omega \big( \sin(\omega \sigma - K) - \sin(\omega \sigma) \big)}
{\omega \sin K - 2 \kappa (1- \cos K)} \,.
\end{split}
\end{align}

\subsection*{Stability of Human-Driven Traffic}
Finally, human-driven traffic can also be analyzed as a special case of ATC or ACC by considering that the ego vehicle that acts like a human driver.
In terms of formulas, this case is recovered by replacing $\alpha$, $\beta$, $\kappa$, $\sigma$ with $\alpha_{\rm H}$, $\beta_{\rm H}$, $\kappa_{\rm H}$, $\tau$.
\subsubsection*{Human-Driven Traffic on a Straight Road}
The straightforward replacement of parameters leads to the plant stability conditions for human drivers in the form
\begin{equation}
\alpha_{\rm H} = 0 \,,
\end{equation}
and
\begin{align}
\begin{split}
\alpha_{\rm H} & = \frac{\Omega^2 \cos(\Omega \tau)}{\kappa_{\rm H}} \,, \\
\beta_{\rm H} & = \Omega \sin(\Omega \tau) - \alpha_{\rm H} \,,
\end{split}
\end{align}
as well as the string stability conditions
\begin{align}
\begin{split}
\alpha_{\rm H} & = 0 \,, \\
\alpha_{\rm H} & = 2(\kappa_{\rm H} - \beta_{\rm H}) \,,
\end{split}
\end{align}
and
\begin{align}
\begin{split}
\alpha_{\rm H} & = \dfrac{\omega^2 \big( \cos(\omega \tau - K) - \cos(\omega \tau) \big)}{\omega \sin K - 2 \kappa_{\rm H} (1 - \cos K)} \,, \\
\beta_{\rm H} & = \dfrac{\omega^2 \cos (\omega \tau) + \kappa_{\rm H} \omega \big( \sin(\omega \tau - K) - \sin (\omega \tau) \big)}{\omega \sin K - 2 \kappa_{\rm H} (1 - \cos K)} \,.
\end{split}
\label{eq:HV_string}
\end{align}

\subsubsection*{Human-Driven Traffic on a Ring}

Human-driven traffic on a ring can be analyzed by the characteristic equation~(\ref{eq:homogeneous_ring_chareq}).
For ${s=0}$ the stability boundary is still
\begin{equation}
\alpha_{\rm H} = 0 \,,
\end{equation}
whereas for ${s={\rm i} \omega}$ it becomes
\begin{align}
\begin{split}
\alpha_{\rm H} & = \dfrac{\omega^2 \big( \cos \big( \omega \tau - \frac{k 2 \pi}{N+1} \big) - \cos(\omega \tau) \big)}{\omega \sin \big( \frac{k 2 \pi}{N+1} \big) - 2 \kappa_{\rm H} \big(1 - \cos \big( \frac{k 2 \pi}{N+1} \big) \big)} \,, \\
\beta_{\rm H} & = \dfrac{\omega^2 \cos (\omega \tau) + \kappa_{\rm H} \omega \big( \sin \big( \omega \tau - \frac{k 2 \pi}{N+1} \big) - \sin (\omega \tau) \big)}{\omega \sin \big( \frac{k 2 \pi}{N+1} \big) - 2 \kappa_{\rm H} \big(1 - \cos \big( \frac{k 2 \pi}{N+1} \big) \big)} \,.
\end{split}
\end{align}
This latter result can be obtained by substituting ${K = 2 k \pi/(N+1)}$ into the string stability boundaries~(\ref{eq:HV_string}) of the human-driven chain of vehicles, as explained in Sec.~\ref{sec:stability_connections}.
Alternatively, the same result can be achieved based on~(\ref{eq:ACC_ring_chareq}, \ref{eq:homogeneous_ring_chareq}) by considering ${\Gamma({\rm i} \omega) = {\rm e}^{{\rm i} \frac{2 k \pi}{N+1}}}$ and substituting
${\Gamma_{\rm R} = \cos \big( \frac{2 k \pi}{N+1} \big)}$, ${\Gamma_{\rm I} = \sin \big( \frac{2 k \pi}{N+1} \big)}$ into~(\ref{eq:ACC_string}) with ${\alpha = \alpha_{\rm H}}$, ${\beta = \beta_{\rm H}}$, ${\kappa = \kappa_{\rm H}}$ and ${\sigma = \tau}$. 

Finally, we remark that the stability of TC, CC followed by human-driven traffic on a straight road or ring, and CC without considering the traffic behind can all be obtained as the special cases of ATC: when there is no response to the distance from the vehicle ahead, i.e., ${\alpha=0}$.
For the ${s=0}$ stability boundaries, however, one needs to take special care since it is exactly ${\alpha=0}$ in ATC.

\bibliographystyle{spmpsci} 
\bibliography{2021_IVSAMA_Molnar}

\end{document}